\documentclass[12pt]{article}
\usepackage[utf8]{inputenc}

\title{AlloyVAE: A generative model for complex probabilistic field-to-field relationships in alloys}

\usepackage{natbib}
\usepackage{amsmath}
\usepackage{epsfig,amsthm}
\usepackage{amssymb,amsfonts}
\usepackage{dsfont}
\usepackage{subfigure}
\usepackage{indentfirst}
\usepackage{multirow} 
\usepackage{booktabs}
\usepackage{mathrsfs}
\usepackage{color}
\usepackage{svg}
\usepackage{xcolor}
\usepackage{graphicx}
\usepackage{zymacros}
\usepackage{hyperref}
\usepackage{algorithm}
\usepackage{authblk}
\usepackage{tikz}
\usepackage{algpseudocode}
\usepackage{amsmath}
\usetikzlibrary{shapes.geometric, arrows}
\begin{document}
\author[1]{Ningyu Yan}
\author[2]{Zhuocheng Xie}
\author[3]{Kai Guo}
\author[3,4,*]{Yejun Gu}
\author[5,*]{Huajian Gao}
\author[1,6,*]{Yang Xiang}
\affil[1]{Department of Mathematics, The Hong Kong University of Science and Technology, Clear Water
Bay, Kowloon, Hong Kong, China}
\affil[2]{Department of Materials Science \& Engineering, University of Toronto, Ontario, Canada M5S 3E4}	
\affil[3]{Institute of High Performance Computing, Agency for Science, Technology and Research, Singapore 138632}
\affil[4]{Department of Mechanical Engineering, Whiting School of Engineering, Johns Hopkins University, Baltimore, MD, USA 21218}
\affil[5]{Mechano-X Institute, Applied Mechanics Laboratory, Department of Engineering Mechanics, Tsinghua University, Beijing, China 100084}
\affil[6]{HKUST Shenzhen–Hong Kong Collaborative Innovation Research Institute, Shenzhen, China}

\maketitle
\noindent\textsuperscript{*}To whom correspondence should be addressed. E-mail: yejun\_gu@a-star.edu.sg (Y.G.), gao.huajian@tsinghua.edu.cn (H.G.), maxiang@ust.hk (Y.X.)
\begin{center}
    \textbf{\large Abstract} \\[1em] 
    \begin{minipage}{0.8\textwidth}
The inherent compositional heterogeneity of multi-principal element alloys (MPEAs) gives rise to complex, spatially varying mechanical fields that cannot be uniquely determined from coarse-grained composition descriptors. This non-uniqueness introduces intrinsically probabilistic structure–property relationships, posing a fundamental challenge to conventional deterministic modeling and machine learning approaches that collapse such mappings into average predictions. Here, we present AlloyVAE, a physics-informed generative framework that learns the full conditional distribution of mechanical fields from microstructural inputs. Built upon a conditional variational autoencoder architecture, the model incorporates learned smoothing operators to enhance functional regularity and a self-consistency mechanism to enforce physical plausibility. Trained on atomistic simulation data, AlloyVAE accurately predicts distributions of residual stress fields from composition and short-range order, and enables the generation of multiple physically consistent realizations under identical input conditions. Beyond forward prediction, the framework supports inverse design by optimizing composition fields to achieve targeted mechanical responses, and is extensible to coupled mappings involving eigenstrain. By capturing one-to-many structure–property relationships in heterogeneous materials, this work establishes a probabilistic paradigm for materials modeling and design, providing a scalable alternative to conventional simulations for navigating high-dimensional compositional spaces.
    \end{minipage}
\end{center}

\section{Introduction}
Multi-principal element alloys (MPEAs) represent a new class of metallic materials in which multiple constituent elements are combined in near-equiatomic proportions~\cite{birbilis2021perspective,coury2021multi,senkov2015accelerated, yeh2004nanostructured}, giving rise to exceptional combinations of strength, ductility, and thermal stability~\cite{lei2018enhanced, li2020complex,an2022new,hung2022thermal}. A defining feature of these alloys is their intrinsic chemical heterogeneity, which induces spatially varying lattice distortions and, consequently, highly heterogeneous strain and stress fields at the nanoscale. Unlike conventional alloys that can often be described by homogenized properties, MPEAs exhibit rich microstructural fluctuations that play a central role in governing deformation, damage initiation, and failure~\cite{varvenne2016theory, varvenne2017solute, geslin2021microelasticity1, geslin2021microelasticity2}.

A fundamental challenge arises when attempting to link such microstructural descriptors to mechanical response. In practice, composition fields and short-range order parameters are obtained through spatial averaging or coarse-graining of underlying atomistic configurations. This process inevitably eliminates detailed configurational information, such that multiple distinct atomic arrangements may correspond to the same coarse-grained description. As a result, the mapping from microstructure to mechanical fields is inherently non-unique: identical composition fields can give rise to different realizations of stress or strain fields, with illustrative examples provided in Section 2 of the Supplementary Information. This intrinsic one-to-many relationship represents a fundamental limitation of existing deterministic modeling frameworks, which implicitly assume a unique correspondence between structure and property.

Recent advances in machine learning have provided powerful tools for accelerating materials modeling and exploring high-dimensional compositional spaces~\cite{wang2023neural,guo2021machine,wen2019machine,wen2021modeling,zhang2021molecular,li2024machine,li2021high}. However, most existing approaches rely on deterministic mappings optimized with mean-squared-error objectives, which collapse the underlying distribution of possible outcomes into a single average prediction. While effective for many regression tasks, such approaches are fundamentally ill-suited to systems where variability is not noise but an intrinsic consequence of coarse-graining. Capturing the full spectrum of physically admissible responses in heterogeneous materials therefore requires a shift from deterministic prediction to probabilistic representation of structure–property relationships~\cite{yang2021deep,yang2021end,yang2023fill}.

Here we introduce AlloyVAE, a physics-informed generative framework designed to learn and sample from the conditional distribution of mechanical fields given microstructural inputs. By integrating a conditional variational autoencoder with learned smoothing operators and a self-consistency mechanism, the framework enables probabilistic field-to-field mapping that preserves both physical conditioning and distributional diversity. This approach allows not only accurate prediction of mechanical field distributions, but also the generation of multiple physically consistent realizations under identical inputs, as well as inverse design of composition fields for targeted mechanical performance. Together, these capabilities establish a new paradigm for modeling heterogeneous materials, in which structure–property relationships are treated as intrinsically probabilistic rather than deterministic.

\section{Results}

\subsection{Alloy variational autoencoder model (AlloyVAE)}
\begin{figure}[htbp]
    \centering
    \includegraphics[scale=0.45]{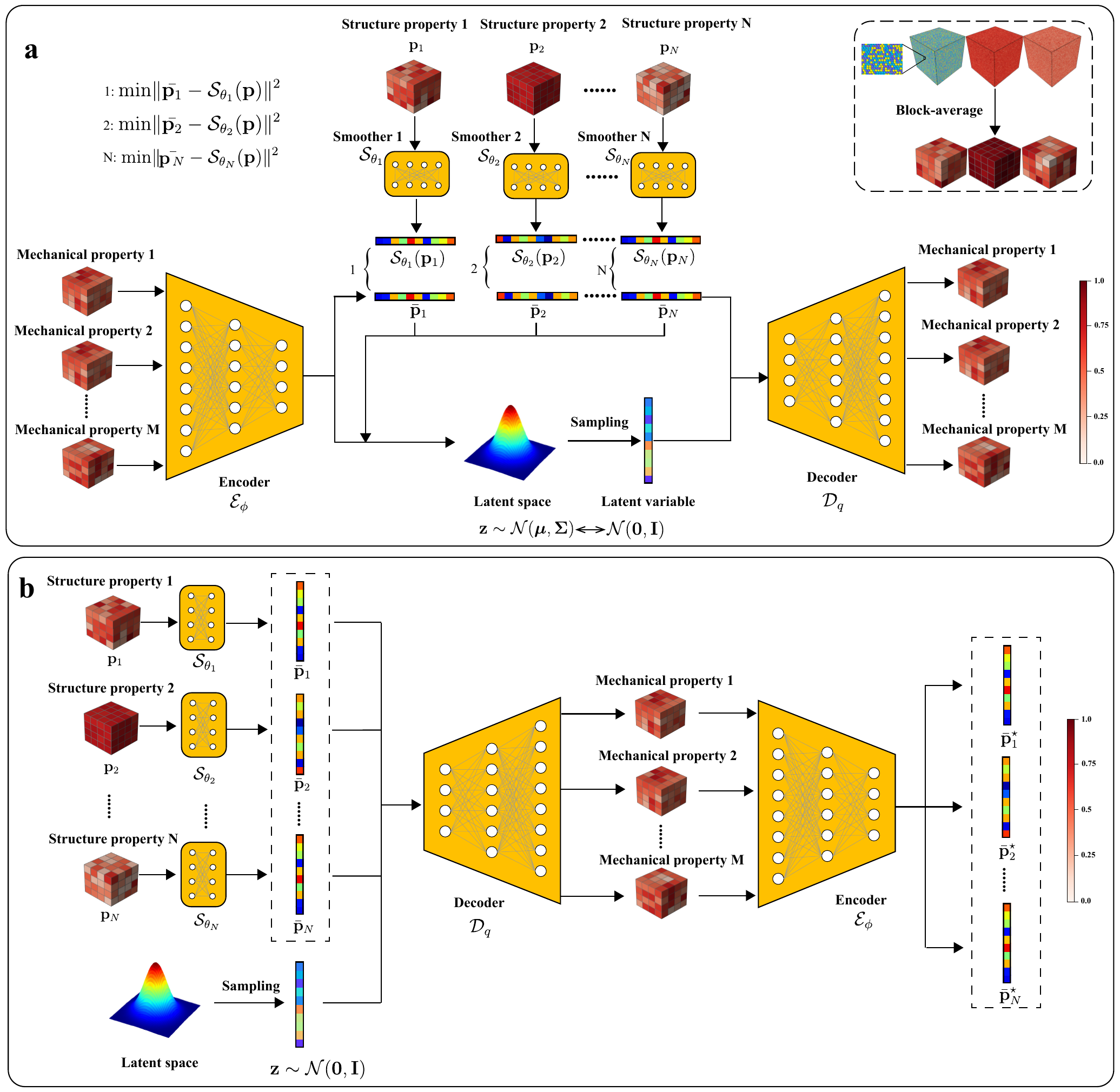}  
    \caption{Architecture of the VAE framework. (a) The general framework used in training phase. The upper right corner shows the block-average process, which transfers the training data from microscale to submicron scale. (b) The general framework used for mechanical property predictions. Both frameworks are modified in model architectures (mainly the smoothers' architectures) and input \& output variables according to the framework in Fig.\ref{VAE framework} to obtain the generalized versions.}  
    \label{general_framework}
\end{figure}

The developed AlloyVAE architecture is designed to perform a probabilistic field-to-field mapping, generating an output field (such as stress) from a given input field (such as composition). It is divided into two distinct components: one for training and one for prediction, as shown in Figure \ref{general_framework}. Each component comprises an encoder, a decoder, and two data smoothers, all implemented through neural networks. During the training phase, the encoder maps the block-averaged (this process is shown in the upper right corner of Figure \ref{general_framework}) target field to a constrained latent distribution. The decoder then samples from this distribution to reconstruct the target field. In the prediction phase, the decoder combines the input field with a randomly sampled latent vector to generate the predicted target field. This architecture enhances VAE generalizability, allowing for the prediction of all plausible target field configurations under specific input field conditions, such as elemental concentrations and Warren–Cowley (WC) parameters quantifying SRO.

The training datasets for AlloyVAE can be collected from experiments or simulations. In this work, the datasets are constructed by uniformly discretizing the simulation cell into sub-blocks and block-averaging the relevant input and target fields for each block. Given that heterogeneous strain caused by lattice distortion at microscopic scales contributes significantly to the enhanced mechanical properties of MPEAs~\cite{li2022heterogeneous}, the length scale of blocks is chosen to span a few nanometers. This length scale is physically critical for capturing the local structural fluctuations that is critical to the mechanical properties. This block-averaging method introduces the challenge of local overfitting, which directly contravenes the stringent regularization requirements of conventional VAEs, and often results in poor training performance. To mitigate this issue, we integrate neural network smoothers into the VAE architecture (see Sec.~\ref{sec:smoother} for a PCA-based analysis of their significance). By imposing a higher regularity within the encoder’s approximation, the smoothers relax the VAE’s sensitivity to atomistic data without compromising its generative capability. Additionally, AlloyVAE employs a conditional VAE (cVAE) framework to address the lack of explicit physical constraints in standard VAE reconstructions. By incorporating the physical input fields as conditioning variables, AlloyVAE is able to explicitly account for spatial variance of the input field and maintain physical consistency across the reconstructed fields.

\begin{figure}[htbp]
    \centering
    \includegraphics[scale=0.45]{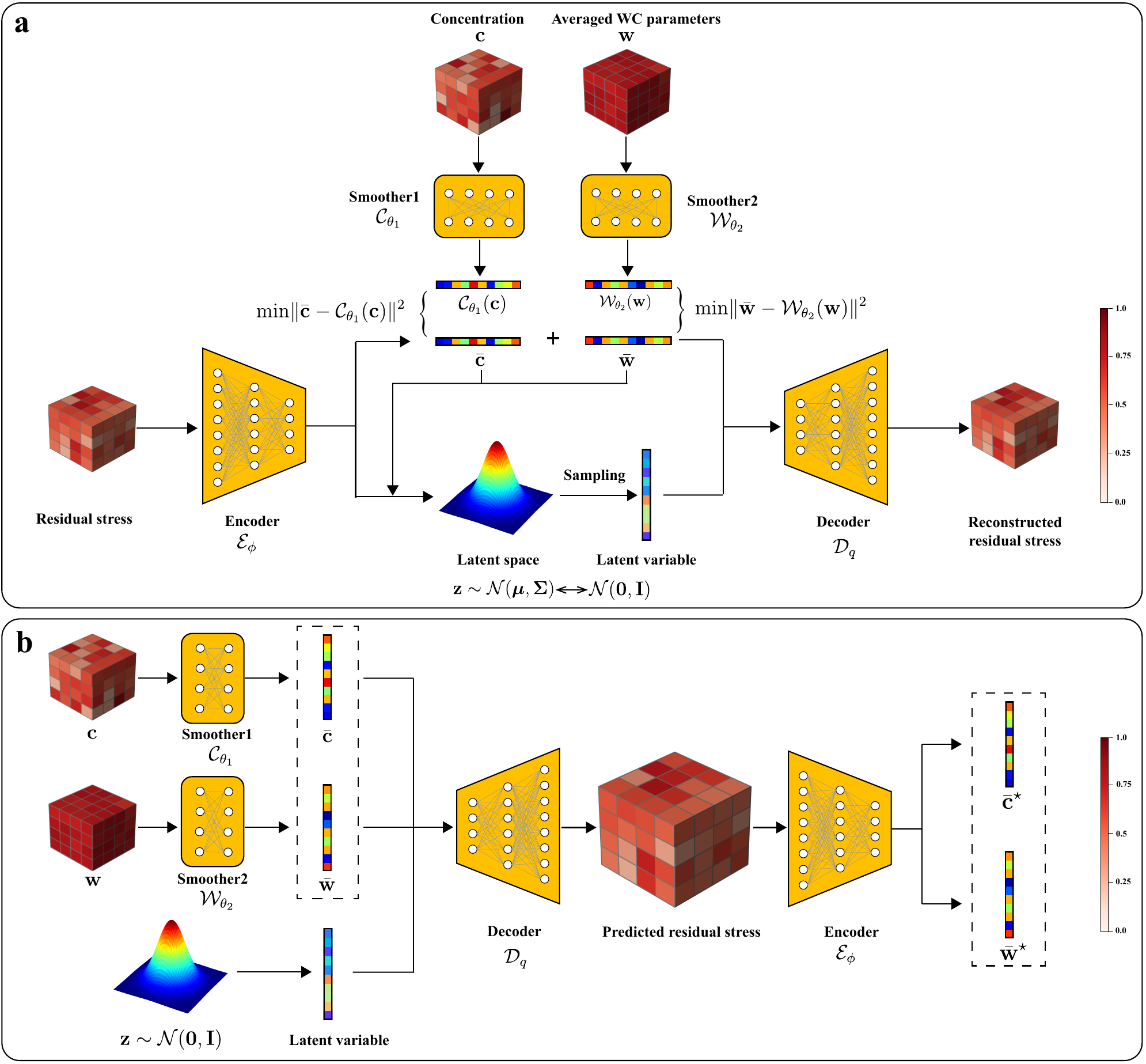}  
    \caption{Architecture of the VAE framework for the prediction from the chemical composition to the residual stress. (a) The framework utilized during training. (b) The framework applied for residual stress prediction. The encoder, $\mathcal{E}_\phi$, receives the residual stress $\boldsymbol{\sigma}$ as input and learns the latent variable $\mathbf{z}$, represented by the mean vector $\boldsymbol{\mu}$ and covariance vector $\boldsymbol{\Sigma}$, along with the smoothed physical conditions $\bar{\mathbf{c}}$ and $\bar{\mathbf{w}}$. The decoder, $\mathcal{D}_q$, reconstructs the residual stress using the latent variable in combination with the physical conditions. Two smoothers, $\mathcal{C}_{\theta_1}$ and $\mathcal{W}_{\theta_2}$, are introduced to enhance the regularity of the function that the encoder approximates.}  

    \label{VAE framework}
\end{figure}

In the following, we will provide more technical details by demonstrating the application of the AlloyVAE for prediction from the chemical composition fields (including the local atomic configurations of the constituent elements and the Warren–Cowley parameters for nine elemental pairs) to the residual stress fields (including six independent components of the virial stress tensors), whose architecture is illustrated in Fig.~\ref{VAE framework}.

\subsubsection{Training phase}
The AlloyVAE framework in the training phase includes four key components, as shown in Fig.\ref{VAE framework}a: \begin{enumerate}
    \item \textbf{Encoder ($\mathcal{E}_{\phi}$)}, which takes stress $\mathbf{\sigma}$ and outputs $\mathbf{\mu}$, $\mathbf{\Sigma}$, $\bar{\mathbf{c}}$, and $\bar{\mathbf{w}}$.
    \item \textbf{Decoder ($\mathcal{D}_{q}$)}, which takes $\mathbf{z}$, $\bar{\mathbf{c}}$, and $\bar{\mathbf{w}}$ and reconstructs $\mathbf{\sigma}$.
    \item \textbf{Latent Space}, in which the latent variable $\mathbf{z}$ is sampled from a constrained distribution $\mathcal{N}(\mathbf{\mu},\mathbf{\Sigma})$.
    \item \textbf{Two Smoothers ($\mathcal{C}_{\theta_1}$ and $\mathcal{W}_{\theta_2}$)}, which transform input fields $\mathbf{c}$ and $\mathbf{w}$ into smoothed conditions $\mathcal{C}_{\theta_1}(\mathbf{c})$ and $\mathcal{W}_{\theta_2}(\mathbf{w})$.
\end{enumerate}
The encoder $\mathcal{E}_\phi$ is designed as a neural network that encodes the residual stress $\boldsymbol{\sigma}$ to its low-dimension space, a $d$-dimension normal latent distribution formulated by mean value vector $\boldsymbol{\mu} \in \mathbb{R}^d $ and a diagonal covariance vector $\boldsymbol{\Sigma}\in \mathbb{R}^d$. Here $\phi$ notes all the parameters in this neural network. The latent variable $\mathbf{z}$, a low-dimension vector representation of residual stress, is randomly sampled from this latent distribution $\mathcal{N}(\boldsymbol{\mu}, \boldsymbol{\Sigma})$, and the reparameterization technique is used for sampling to allow gradient backpropagation in neural networks: 
\begin{equation}
\mathbf{z}=\boldsymbol{\mu}+\boldsymbol{\varepsilon}\odot \boldsymbol{\Sigma}, 
\end{equation}
where $\odot$ is the element-wise multiplication operator, and the noise $\boldsymbol{\varepsilon}$ is randomly sampled from $\mathcal{N}(\mathbf{0},\mathbf{I})$. Additionally, to enforce consistency with the composition field, the encoder is designed to also predict physical conditions ($\bar{\mathbf{c}}$ and $\bar{\mathbf{w}}$) as auxiliary outputs. Two smoothers, neural networks $\mathcal{C}_{\theta_1}$ and $\mathcal{W}_{\theta_2}$, approximate these physical conditions by transforming the raw block-averaged concentration field $\mathbf{c}$ and the short-range order field $\mathbf{w}$ to $\mathcal{C}_{\theta_1}(\mathbf{c})$ and $\mathcal{W}_{\theta_2}(\mathbf{w})$, respectively. These smoothers serve to mitigate local overfitting caused by block-averaging and enhance the regularity of the function the encoder approximates. Here, $\theta_1$ and $\theta_2$ denote the parameters in the two neural networks. By minimizing the distance between the physical conditions predicted by both the encoder and the smoothers, we effectively integrate the smoothed composition field to accelerate convergence and stabilize the training process. Finally, the decoder $\mathcal{D}_q$ is a neural network that utilizes the sampled vector $\mathbf{z}$ along with the physical conditions as inputs to reconstruct the residual stress $\boldsymbol{\sigma}$. Here $q$ notes all the parameters in this neural network. 


The goal of the training phase is to optimize the AlloyVAE framework to learn the distribution of residual stress given the input of chemical composition fields. This is achieved by maximizing the evidence lower bound (ELBO) of the likelihood function, conditioned on the composition fields. Given a dataset $D=\{(\mathbf{c}_i, \mathbf{w}_i, \boldsymbol{\sigma}_i) \mid i=1,2,\dots,N\}$ consisting of $N$ data samples, the encoder, decoder, and two smoothers can be jointly trained to optimize the hyperparameter set $\phi, q, \theta_1$, and $\theta_2$ with the overall loss function, $\mathcal{L}$. The loss function is formulated as 
\begin{align}
    \mathcal{L}= &\underbrace{\sum_{i=1}^{N} D_{KL}\big(\mathcal{N}(\boldsymbol{\mu}_i, \diag(\boldsymbol{\Sigma}_i))\|
    \mathcal{N}(\mathbf{0},\mathbf{I})\big)}_{\text{KL divergence loss for for encoder}}+\underbrace{\sum_{i=1}^{N}\|\mathcal{D}_q(\mathbf{z}_i,\mathbf{c}_i, \mathbf{w}_i)-\boldsymbol{\sigma}_i \|^2}_{\text{MSE loss for decoder}}\nonumber\\
    &+\underbrace{\underbrace{\alpha\sum_{i=1}^{N}\|\bar{\mathbf{c}}_i-\mathcal{C}_{\theta_1}(\mathbf{c}_i) \|^2}_{\text{concentration field condition}}+\underbrace{\beta\sum_{i=1}^{N}\|\bar{\mathbf{w}}_i-\mathcal{W}_{\theta_2}(\mathbf{w}_i) \|^2}_{\text{SRO field condition}}}_{\text{MSE loss for physical input condition}}.
    \label{loss}
\end{align}
The first term of Eq.~\eqref{loss}, the KL divergence loss, originates from the assumption that we aim to generate samples with the input of latent variables sampled from $\mathcal{N}(\mathbf{0},\mathbf{I})$\cite{kingma2013auto}, and it aims to minimize the distance between the latent
space distribution $\mathcal{N}(\boldsymbol{\mu}_i, \diag(\boldsymbol{\Sigma}_i)$ generated from the encoder and $\mathcal{N}(\mathbf{0},\mathbf{I})$, which can be simplified to 
\begin{equation}
    D_{KL}\big(\mathcal{N}(\boldsymbol{\mu}_i, \diag(\boldsymbol{\Sigma_i}))\|
    \mathcal{N}(\mathbf{0},\mathbf{I})\big)=\sum_{j=1}^{d} \frac{1}{2}\left(-1+(\Sigma_i^j)^2+(\mu_i^j)^2-\log{(\Sigma_i^j)^2}\right).
\end{equation}
The second term of Eq.~\eqref{loss} minimizes the mean squared error (MSE) between the reconstruction results $\mathcal{D}_q(\mathbf{z}_i,\mathbf{c}_i, \mathbf{w}_i)$ by the decoder and reference. These two terms constitute the negative ELBO of the log-likelihood for the residual stress distribution conditioned on the composition field. 
The last two terms of Eq.~\eqref{loss} penalize the prediction errors of the physical conditions,  enforcing consistency between the physical conditions predicted by the encoder ($\bar{\mathbf{c}}$ and $\bar{\mathbf{w}}$) and the ones by the smoothers ($\mathcal{C}_{\theta_1}(\mathbf{c})$ and $\mathcal{W}_{\theta_2}(\mathbf{w}$)), respectively. Here, $\alpha$ and $\beta$ are adjustable weights and $\bar{\mathbf{c}}_i,\bar{\mathbf{w}}_i, \boldsymbol{\mu}_i$,  $\boldsymbol{\Sigma}_i$ are the outputs of $\mathcal{E}_\phi(\boldsymbol{\sigma}_i)$. 

\subsubsection{Prediction phase}
After training the framework, the VAE is used to predict residual stress $\boldsymbol{\sigma}$ by leveraging the learned latent space. The concentration, $\mathbf{c}$, and average WC parameters, $\mathbf{w}$, are first processed through two smoothers to obtain the smoothed conditions, $\bar{\mathbf{c}}$ and $\bar{\mathbf{w}}$, respectively. To account for the high variance in the deep learning model’s output, which may arise due to the block-averaged dataset, a latent variable $\mathbf{z}$ is randomly sampled from $\mathcal{N}(\mathbf{0},\mathbf{I})$. This latent variable, along with $\bar{\mathbf{c}}$ and $\bar{\mathbf{w}}$, is then fed into the decoder to predict the residual stress, $\boldsymbol{\sigma}$.
However, due to the discrepancy between the learned latent space $\mathcal{N}(\boldsymbol{\mu}_i, \text{diag}(\boldsymbol{\Sigma}_i))$ and the standard Gaussian prior $\mathcal{N}(\mathbf{0},\mathbf{I})$, the randomly sampled $\mathbf{z}$ may not always yield a physically reasonable residual stress prediction. As a result, when the predicted residual stress, $\boldsymbol{\sigma}$, is fed back into the encoder to reconstruct the physical conditions, i.e., the concentration $\bar{\mathbf{c}}^\star$ and the WC parameters $\bar{\mathbf{w}}^\star$, The reconstructed values $\bar{\mathbf{c}}^\star$ and $\bar{\mathbf{w}}^\star$ may differ significantly from the input of physical conditions $\bar{\mathbf{c}},\bar{\mathbf{w}}$. 
To address this, the plausibility of the predicted residual stress is assessed using a self-checking algorithm: The predicted residual stress is accepted only if the distance between the reconstructed physical conditions and the input of smoothed physical conditions is sufficiently small, i.e., does not exceed a predefined threshold, $k$. Otherwise, a new latent variable is sampled, and the stress prediction process is repeated until the physical condition criterion is satisfied, the required number of valid residual stress predictions $N_g$ is achieved, or the maximum iteration limit $N$ is reached.

By incorporating the selected latent variables with physical conditions, the model produces reliable residual stress predictions while maintaining consistency between the reconstructed and actual physical conditions, ensuring that discrepancies remain within a predefined threshold. All different but self-checking guaranteed residual stress are regarded as reasonable predictions corresponding to the input concentration and averaged WC parameters, which can serve as valid reference for practical applications, such as reducing the design space and trial-error costs significantly. The prediction process is illustrated in Algorithm~\ref{a1}.

\begin{algorithm}
\caption{Residual Stress Prediction Process with Self-checking}
\label{a1}
\begin{algorithmic}[1]
\State \textbf{Input:} $\mathbf{c}, \mathbf{w}, \mathcal{C}_{\theta_1}, \mathcal{W}_{\theta_2}, \mathcal{D}_q, N, N_g, k, n = 0$
\State \textbf{Calculate:} $\bar{\mathbf{c}} = \mathcal{C}_{\theta_1}(c)$
\State \textbf{Calculate:} $\bar{\mathbf{w}} = \mathcal{W}_{\theta_2}(w)$
\For{$i = 1, 2, \dots, N$}
    \State Sample a latent variable $\mathbf{z} \sim \mathcal{N}(0, I)$
    \State Predict residual stress: $\boldsymbol{\sigma} = \mathcal{D}_q(\bar{\mathbf{c}}, \bar{\mathbf{w}}, \mathbf{z})$
    \State Reinput $\boldsymbol{\sigma}$ to encoder: $\mathcal{E}_{\phi}(\boldsymbol{\sigma}) \to \bar{\mathbf{c}}^*, \bar{\mathbf{w}}^*$
    \If{MRE($\bar{\mathbf{c}}^*, \bar{\mathbf{c}}$) $\leqslant k$ and MRE($\bar{\mathbf{w}}^*, \bar{\mathbf{w}}$) $\leqslant k$}
        \State Output residual stress $\boldsymbol{\sigma}$
        \State $n = n + 1$
    \EndIf
    \If{$n > N_g$}
        \State \textbf{break}
    \EndIf
\EndFor
\end{algorithmic}
\end{algorithm}
\subsubsection{The importance of smoothers}
\label{sec:smoother}
\begin{figure}[htbp]
    \centering
    \includegraphics[scale=0.3]{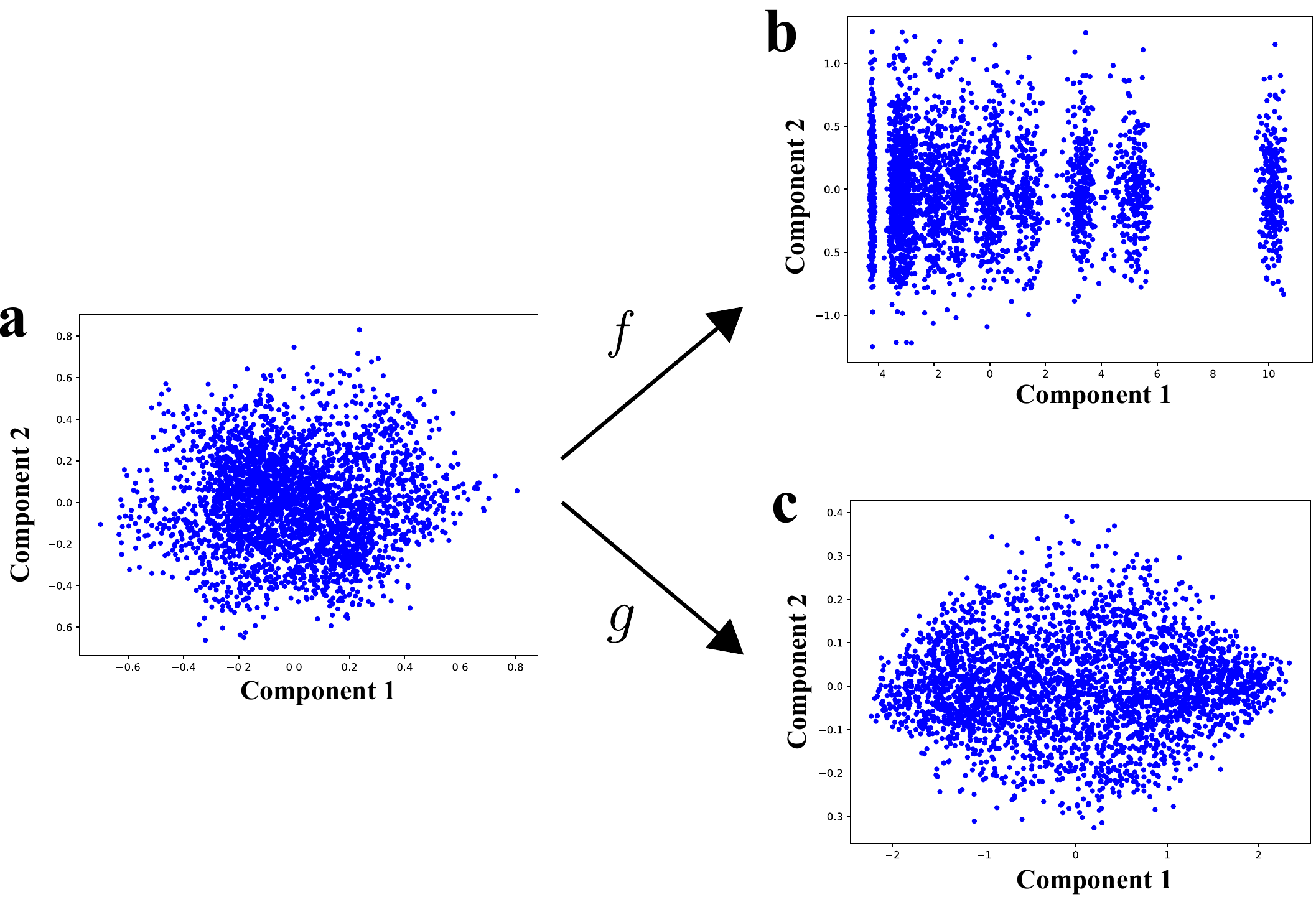}  
    \caption{PCA visualization. $f$, $g$ are the highly non-linear functions from residual stress $\boldsymbol{\sigma}$ to the original features ($\mathbf{c}$, $\mathbf{w}$) and physical conditions ($\bar{\mathbf{c}}$, $\bar{\mathbf{w}}$), respectively. Component 1 and Component 2 are two principal components of PCA. This is equivalent to $f$ being the function that the encoder needs to approximate without smother, and $g$ being the function that needs to be approximated with smother. a) The PCA result of residual stress samples in train dataset. b) The PCA result of merging samples of concentration $\mathbf{c}$ and WC parameters $\mathbf{w}$. c) The PCA result of merging samples of smoothed concentration $\bar{\mathbf{c}}$ and smoothed WC parameters $\bar{\mathbf{w}}$. }
    \label{PCA visualization}
\end{figure}

As compared to the conventional VAE, we have incorporated two smoothers ($\mathcal{C}{\theta_1}$ and $\mathcal{W}{\theta_2}$) to address the poor reconstruction performance associated with raw concentration and SRO parameter data. To justify the role of the smoothers, principal component analysis (PCA) is applied to project the residual stress, the original feature pair ($\mathbf{c}$, $\mathbf{w}$), and the smoothed physical conditions ($\bar{\mathbf{c}}$, $\bar{\mathbf{w}}$) from the training dataset into a two-dimensional space, facilitating clearer visualization of their respective data distributions. As illustrated in Fig.~\ref{PCA visualization}, the raw features ($\mathbf{c}$, $\mathbf{w}$), obtained through the mapping $f$ from residual stress to the original feature space, exhibit a striped distribution. In contrast, the mapping $g$, formed by coupling $f$ with the smoothers, produces more homogeneous and uniformly distributed representations of the smoothed features ($\bar{\mathbf{c}}$, $\bar{\mathbf{w}}$) and residual stress. The PCA results suggest that the function $g$ possesses greater regularity than $f$. The striped and clustered structure of the raw feature space can introduce discontinuities, where the left and right limits of the mapping $f$ may not coincide. This indicates the potential non-differentiability in $f$ at certain points and presents a significant challenge for neural network approximation, which fundamentally depends on the smoothness of target functions~\cite{yarotsky2017error, yarotsky2018optimal, petersen2018optimal, de2021approximation, langer2021approximating}. Specifically, Langer et al.~\cite{langer2021approximating} proved that for neural networks with sigmoid activation functions, as used in our implementation, and employing $M^{2d}$ weights, the approximation error can be bounded by $O(M^{-2p})$, where $p$ denotes the degree of regularity of the target function,
\begin{equation} \sup_{\mathbf{x} \in [-a,a]^d} |f_{NN}(x)-f(x)| \leqslant C\cdot M^{-2p}, \end{equation} 
where $C$ is a finite constant, and $f_{NN}$ is the approximation from the neural networks of a target function $f: \mathbb{R}^d \to \mathbb{R}$. Therefore, the two smoothers enhance the regularity of the target function, contributing to the high predictive accuracy of the model. Moreover, they significantly reduce the dependence on randomness and the representativeness of the training data during the data generation process, thereby improving the generalization capability of AlloyVAE even when trained on data that is not fully representative of the underlying distribution.


\subsection{Performance of AlloyVAE}
In this work, CoCrNi is selected as a representative multi-principal element alloy (MPEA) due to its promising application potential and extensive research coverage~\cite{wu2014temperature, miao2017evolution, george2019high, zhang2020short, liu2022exceptional}. Atomistic simulations were conducted using the Large-scale Atomic/Molecular Massively Parallel Simulator (LAMMPS)~\cite{LAMMPS} to compute the virial stress field under load-free conditions with periodic boundary constraints, and the results were converted to spatially averaged values through the block-average method (see details in the Methods Section 4.1.1). The primary objective is to ensure the robustness of the AlloyVAE framework, which requires high reconstruction accuracy and strong predictive performance from the encoder. To assess reconstruction performance, residual stress from the test dataset is fed into the encoder. The decoder then reconstructs the residual stress using the sampled latent variables and physical conditions. 

\begin{figure}[htbp]
    \centering
    \includegraphics[scale=0.45]{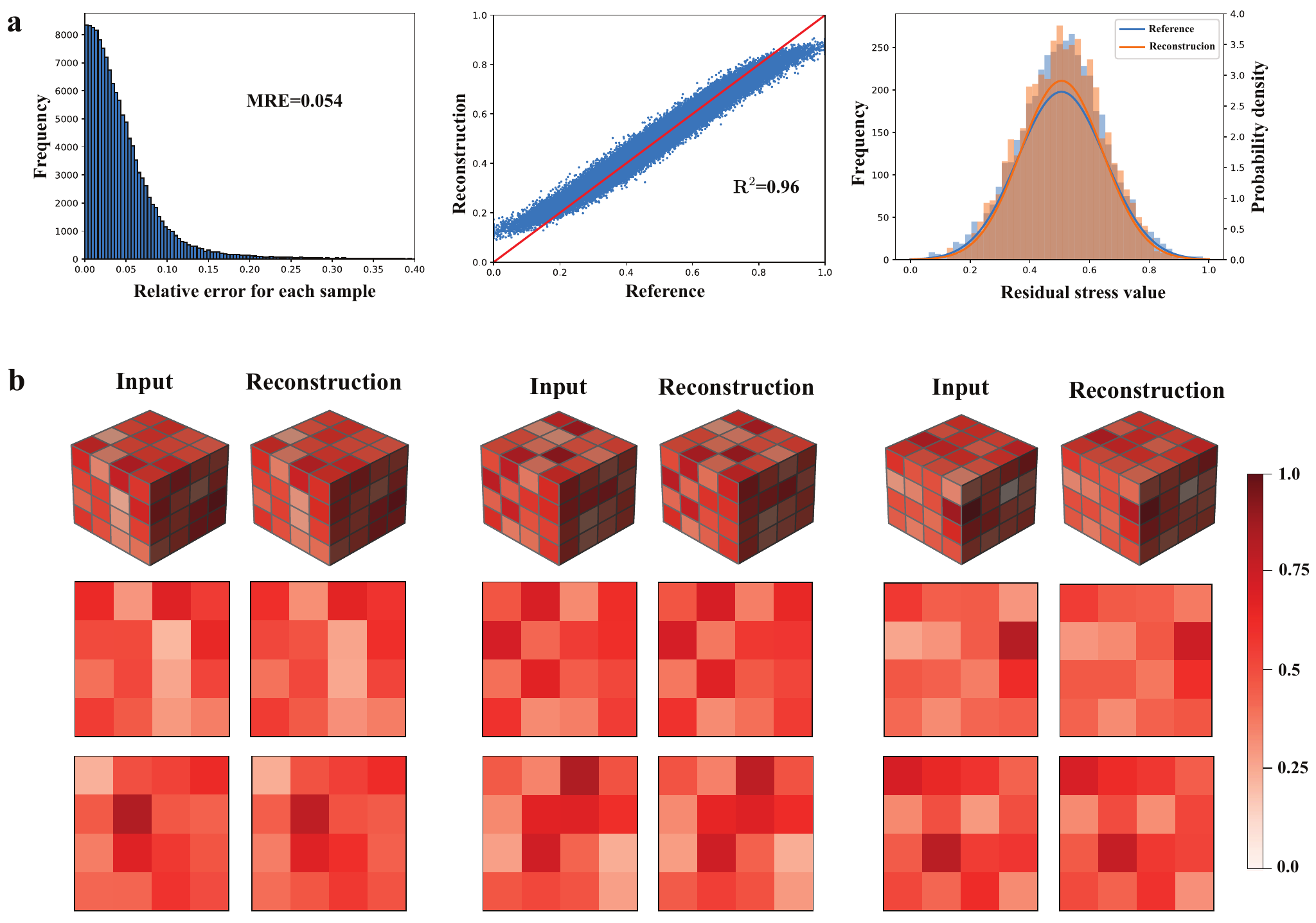}  
    \caption{VAE reconstruction performance. a) The left is the histogram of the frequency distribution of relative errors for all samples in the test dataset, and the MRE of dataset is 5.4\%. The middle is the reconstruction vs. reference of each residual stress block value, the red solid line has a zero intercept and unity slope; the $\mathbf{R}^2$ score is 0.96. The right is the histogram of all residual stress block value of reconstruction and reference, both of them exhibit the tendency toward normal distribution. b) Three reconstruction vs. input reference examples of the residual stress for the \(xx-\)component. The first row shows the reconstruction result at block level, the following 2 rows show the reconstruction at top 2 cross sections along the x-axis. }
    \label{VAE reconstruction performance}
\end{figure}

The predictive accuracy of AlloyVAE is quantified by the mean relative error (MRE) of the stress component $\sigma$, defined as the average deviation between the ground-truth and reconstructed fields across all blocks of all MD simulation samples:  
\begin{equation}
    \mathrm{MRE}=\frac{1}{N_t\times r^3}\sum_{i=1}^{N_t} \sum_{j=1}^{r^3} \frac{|\sigma^{rec}_{ij}-\sigma^{ref}_{ij}|}{|\sigma^{ref}_{ij}|},
\end{equation}
where $N_t$ denotes the total number of MD simulation samples, and $r$ represents the number of blocks along each dimension of the samples. MRE provides a normalized metric of the reconstruction accuracy across the dataset. Due to sample size constraints in atomistic simulations, we set $r=4$ as a coarse yet practical choice, corresponding to a block size of approximately 7.5 nm. The residual stress component, $\sigma_{ij}$, is defined with $i$ as the sample index and $j$ as the block index within a sample. Fig.\ref{VAE reconstruction performance}a shows the comparison between the reconstruction results and the ground truth data for the \(xx-\)component of residual stress. Fig.~\ref{VAE reconstruction performance}b illustrates the frequency distribution histogram of all reconstruction and reference datapoints, and we also calculate the $R^2$ in Fig. \ref{VAE reconstruction performance}c with $R^2= 95\%$. All these results demonstrate that the VAE framework accurately reconstructs residual stress while maintaining a high degree of fidelity in capturing fine-scale local features.

\begin{figure}[htbp]
    \centering
    \includegraphics[scale=0.53]{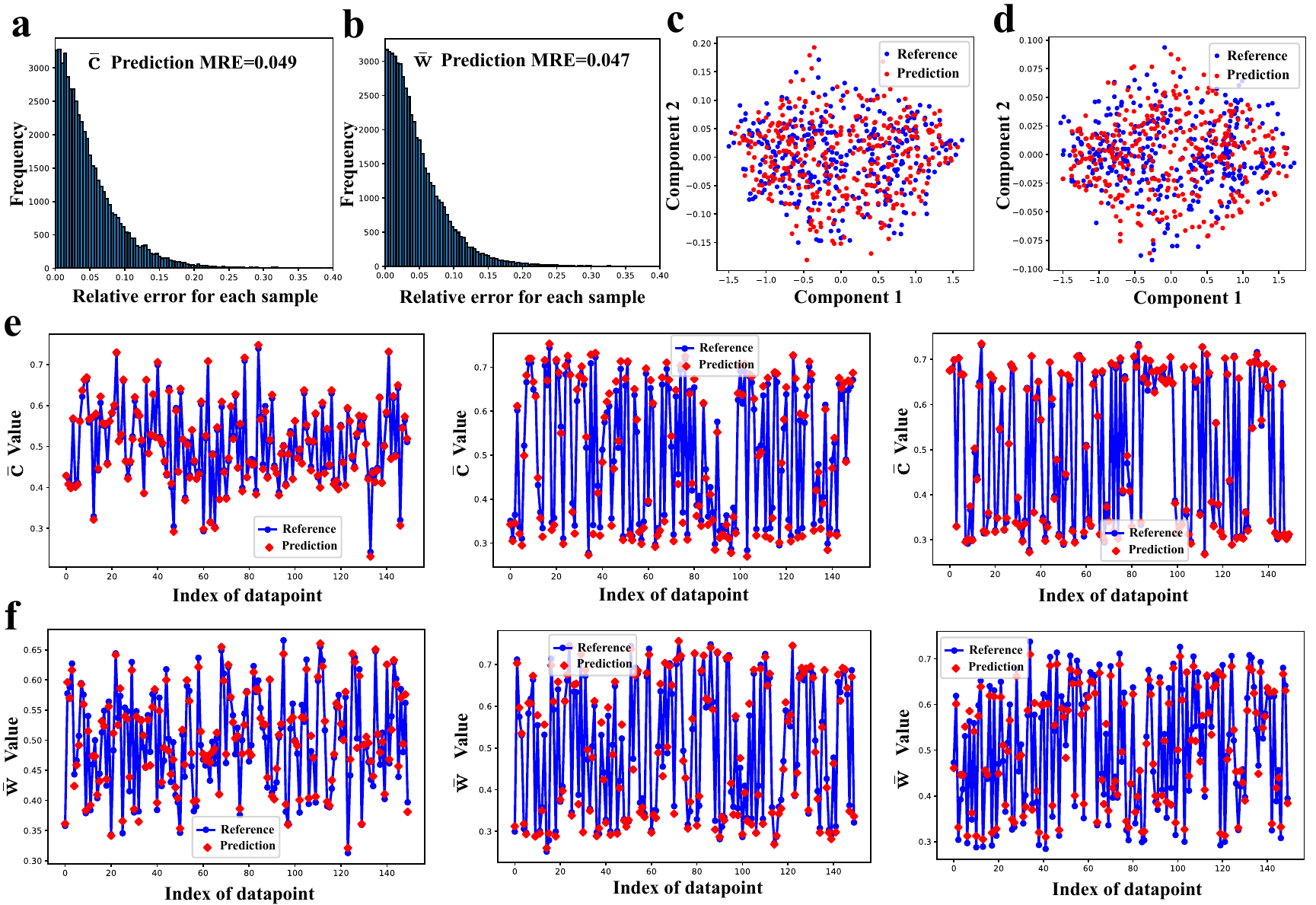}  
    \caption{VAE encoder performance. (a, b) Histograms of relative errors for all test samples under physical conditions $\bar{\mathbf{c}}$ and $\bar{\mathbf{w}}$, respectively. (c, d) PCA projections of the test samples for $\bar{\mathbf{c}}$ and $\bar{\mathbf{w}}$, respectively. (e, f) Examples of predicted values (red dots) versus reference values (blue solid lines) under selected physical conditions.}
    \label{VAE smoothers performance}
\end{figure}

Subsequently, we evaluate the encoder's performance in predicting the physical conditions. The residual stress from the test dataset is fed into the encoder to predict the physical conditions \(\bar{\mathbf{c}}, \bar{\mathbf{w}}\), while the corresponding \(\mathbf{c}, \mathbf{w}\) are input into smoothers to obtain the reference conditions \(\mathcal{C}_{\theta_1}(\mathbf{c}), \mathcal{W}_{\theta_2}(\mathbf{w})\). The MRE is then computed between the predicted and reference values. As shown in Fig.~\ref{VAE smoothers performance}a,b, the MREs between \(\bar{\mathbf{c}}, \bar{\mathbf{w}}\) and \(\mathcal{C}_{\theta_1}(\mathbf{c}), \mathcal{W}_{\theta_2}(\mathbf{w})\) are both below 4.9\%, indicating a high level of predictive accuracy. Furthermore, principal component analysis (PCA) is used to visualize the 2D representations of the predicted and reference physical conditions (concentration and SRO parameters in Fig.~\ref{VAE smoothers performance}c,d, respectively), demonstrating strong consistency of the distribution in the reduced-dimensional space. 

Furthermore, Fig.~\ref{VAE smoothers performance}e,f present three representative examples (three columns) of the predicted physical conditions. Beyond the strong alignment between the predictions and references, it is noteworthy that the learned physical conditions are not just constant-valued points but exhibit oscillatory behavior. This suggests that the predicted physical conditions \(\bar{\mathbf{c}}, \bar{\mathbf{w}}\) effectively capture the intrinsic characteristics of \(\mathbf{c}, \mathbf{w}\), ensuring the reliability of the guidance they provide. 

Thus, these results confirm the effectiveness of the VAE framework, demonstrating its capability for future prediction of residual stress. 


\subsection{Residual stress prediction}
\begin{figure}[htbp]
    \centering
    \includegraphics[scale=0.34]{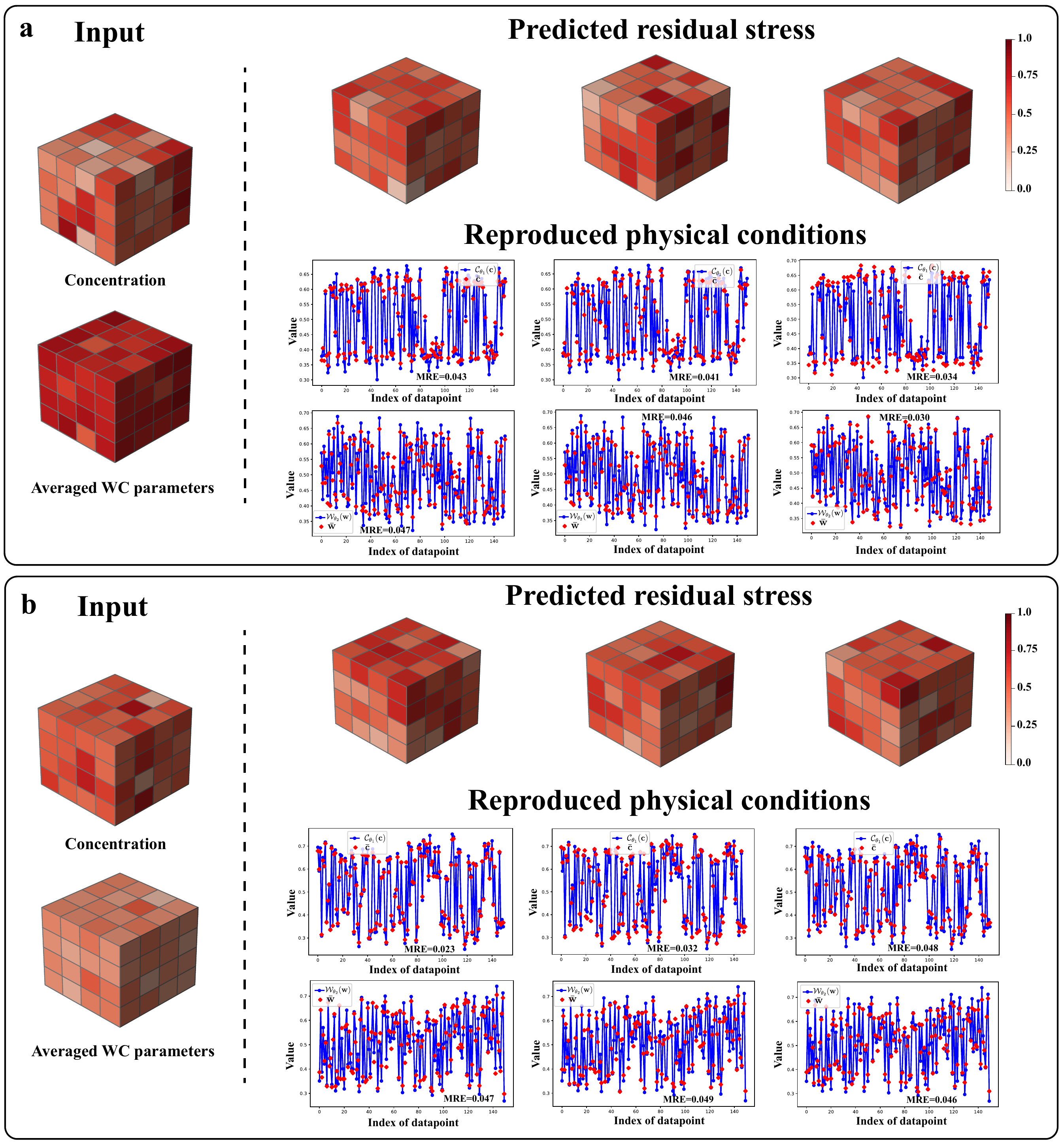}  
    \caption{Residual stress prediction through self-checking process. For each subfigure, the left is the reference residual stress in test dataset. 3 reasonable  predicted residual stress (only show the \(xx-\)component) with their reproduced physical conditions are shown on the right. All resulting MRE values remain below 4\%.}
    \label{Residual_stress_prediction}
\end{figure}

The residual stress prediction process is executed using Algorithm\ref{a1}, which incorporates a self-checking mechanism. We establish the maximum number of iterations at $N=500$ and set the validation threshold $k$ to 4\%. The objective is to generate $N_g=3$ physically plausible residual stress distributions for each input defined by the concentration and WC parameters. Fig.~\ref{Residual_stress_prediction} presents two illustrative examples of residual stress prediction, where each example yields three distinct scenarios of predicted residual stress along with the reproduced physical conditions. While variability exists among the predicted stress fields, the capacity to accurately reproduce physical conditions (evidenced by the MRE being below the $4\%$ threshold) validates the precision and rationality of the predictions when coupled with the self-checking procedure. All scenarios satisfying the prescribed threshold represent rational predictions, given the inherent generalization error induced by the block-averaging of the dataset. This comprehensive set of results provides a valuable reference for further continuum-scale real-world applications, such as defect detection and strength assessment. Furthermore, the requisite input fields, $\mathbf{c}$ and $\mathbf{w}$, can be empirically measured by techniques such as scanning electron microscopy (SEM)\cite{zhou2024wear} and atom probe tomography (APT)\cite{devaraj2024quantifying}. The comparison of the MSE-based ML model and probabilistic generative models in capturing those one-to-many mappings is detailed in the SI.


\subsection{Strengthening through concentration optimization}
\label{sec:strengthening}
In materials design, optimizing mechanical performance requires precise control over elemental concentration distributions. In this work, we demonstrate the application of the proposed AlloyVAE model to design an equiatomic CoCrNi alloy that exhibits enhanced dislocation resistance. 

Without loss of generality, we assume isotropic conditions and consider a single slip system characterized by the Burgers vector $\mathbf{b} = \frac{1}{\sqrt{2}}(1, -1, 0)$ and the slip plane normal $\mathbf{n} = \frac{1}{\sqrt{3}}(1, 1, 1)$ in a purely random equiatomic CoCrNi system (with all WC parameters set to zero), the corresponding resolved shear stress is given by
\begin{equation}
    \tau_p = \mathbf{n}\cdot \boldsymbol{\sigma}\cdot \frac{\mathbf{b}}{|\mathbf{b}|}=\frac{1}{\sqrt{6}}\big(\sigma_{11}-\sigma_{22}+\sigma_{13}-\sigma_{23}\big),
\end{equation}
where $\cdot$ denotes the single-index contraction, adopting the standard convention for vector-tensor products. Thus $\tau_p$ is inherently dependent on concentration and the WC parameter. 
Rida et al \cite{rida2022influence} have formulated that the characteristic resistance for dislocation slip, $\sigma_\tau$, can be scaled by the variance of the resolved shear stress, $\langle\tau_p^2\rangle$, i.e., $\sigma_\tau \sim \langle\tau_p^2\rangle$. Therefore, strength enhancement in the context of materials design can be formulated as a concentration optimization problem aimed at maximizing the variance of the resolved shear stress, $\langle \tau_p^2 \rangle$, which is evaluated using AlloyVAE. The optimization is constrained by three conditions: equiatomic alloy composition, random alloy selection due to the technical difficulty in precisely controlling the SRO during processing, and a self-check for concentration and WC parameters to ensure the validity of residual stress predictions. Consequently, the formulation of the optimization problem is
\begin{equation}
    \underset{\mathbf{c},\mathbf{z}}{\operatorname*{\operatorname*{\arg\min}}}f(\mathbf{c},\mathbf{z}),
    \label{op2}
\end{equation}
 where the objective function  $f$ is
\begin{equation}
f(\mathbf{c},\mathbf{z})=
    \alpha \frac{1}{\langle\tau_p^2\rangle}+\xi\sum_{i=1}^{3} (c_i-c_{\mathrm{equi}})^2+\beta\|\bar{\mathbf{c}}-\bar{\mathbf{c}}^\star\|^2+\gamma\|\bar{\mathbf{w}}-\bar{\mathbf{w}}^\star\|^2,
    \label{op}
\end{equation}
where $\alpha,\xi,\beta,\gamma$ are adjustable weights, $\bar{\mathbf{c}}^\star, \bar{\mathbf{w}}^\star$ are reproduced physical conditions by the encoder $\mathcal{E}_\phi$, $c_i$ is the concentration of $i-$th alloy element. For the equiatomic random samples with no SRO, the normalized equiatomic concentration, $c_\mathrm{equi}$, and all components of the average WC parameter, $\mathbf{w}$, are set to a value of 0.5, which corresponds to the pre-normalization states of $1/3$ and $0$, respectively. 
Here, optimization is performed with respect to concentration $\mathbf{c}$ together with latent variable $\mathbf{z}$.
 The first term of $f$ is designed to maximize the variance of the resolved shear stress $\tau_p$, the second term ensures that the equiatomic condition is satisfied, and the last two terms enforce the self-checking process on the concentration $\bar{\mathbf{c}}$ and the WC parameters $\bar{\mathbf{w}}$. The residual stress $\boldsymbol{\sigma}$, along with the concentration $\bar{\mathbf{c}}^\star$ and the WC parameters $\bar{\mathbf{w}}^\star$, in the objective function $f$ (Eq.~\eqref{op}) are computed by a well-trained AlloyVAE with all values of the inputs $\mathbf{c}, \mathbf{z}$ and the WC parameters $\mathbf{w}$ equal to $0.5$.

This optimization problem can be solved by a gradient descent method, in which the gradient of $f$ with respect to $\mathbf{c}$ and $\mathbf{z}$ is calculated  using automatic differentiation \cite{baydin2018automatic}, and a learning rate $\eta$, fixed at $10^{-2}$, is used to update $\mathbf{c}$ and $\mathbf{z}$ in the gradient descent. 
The entire optimization process is summarized in Figure \ref{op_concen} (a) and Algorithm \ref{op_algo}.
\begin{algorithm}
\caption{Concentration Optimization Process}
\label{op_algo}
\begin{algorithmic}[1]
\State \textbf{Initialization:} $\mathbf{c}^{(0)}, \mathbf{z}^{(0)}$, fixed $\mathbf{w}$, $\eta$
\For{$t \leftarrow $ 1 \textbf{to} $N$}
    \State Calculate physical conditions: $\bar{\mathbf{c}}^{(t-1)}=\mathcal{C}_{\theta_1}(\mathbf{c}^{(t-1)})$, $\bar{\mathbf{w}}=\mathcal{W}_{\theta_2}(\mathbf{w})$
    \State Predict residual stress: $\boldsymbol{\sigma}^{(t-1)} = \mathcal{D}_q(\bar{\mathbf{c}}^{(t-1)}, \bar{\mathbf{w}}, \mathbf{z}^{(t-1)})$
    \State Reconstruct physical conditions: 
$\bar{\mathbf{c}}^{\star(t-1)},\bar{\mathbf{w}}^{\star(t-1)} = \mathcal{E}_{\phi}(\boldsymbol{\sigma}^{(t-1)})$
    \State Evaluate $f(\mathbf{c}^{(t-1)}, \mathbf{z}^{(t-1)})$ in equation (\ref{op})
    \State Calculate gradients: $\nabla_{\mathbf{c}^{(t-1)}}f, \nabla_{\mathbf{z}^{(t-1)}}f$ 
    \State Update $\mathbf{c}^{(t)}=\mathbf{c}^{(t-1)}-\eta\times\nabla_{\mathbf{c}^{(t-1)}}f$
    \State Update $\mathbf{z}^{(t)}=\mathbf{z}^{(t-1)}-\eta\times\nabla_{\mathbf{z}^{(t-1)}}f$
\EndFor
\State \textbf{Return:} $\mathbf{c}^{(N)}$
\end{algorithmic}
\end{algorithm}

To evaluate the transferability of the proposed method across different datasets, we initialize the optimization problem using 2000 concentration distributions randomly sampled from a normal distribution, whose mean and variance are derived from the training dataset. This approach prevents direct reuse of concentration profiles from the training data. 

For the initialization of latent variables, we consider two strategies: one is sampling 2000 initial latent vectors from a standard normal distribution $\mathcal{N}(\mathbf{0}, \mathbf{I})$ (referred to as the Random Latent Method); and the other is assigning a fixed latent variable to each initial concentration distribution (referred to as the Fixed Latent Method). The optimization is performed over 1000 iterations using hyperparameters $\alpha = 0.1$, $\beta = \gamma = 50$, and $\xi = 1$. Each pair of initial concentration and latent variable is then passed to the optimization algorithm to obtain the optimized concentration.

\begin{figure}[htbp]
    \centering
    \includegraphics[scale=0.4]{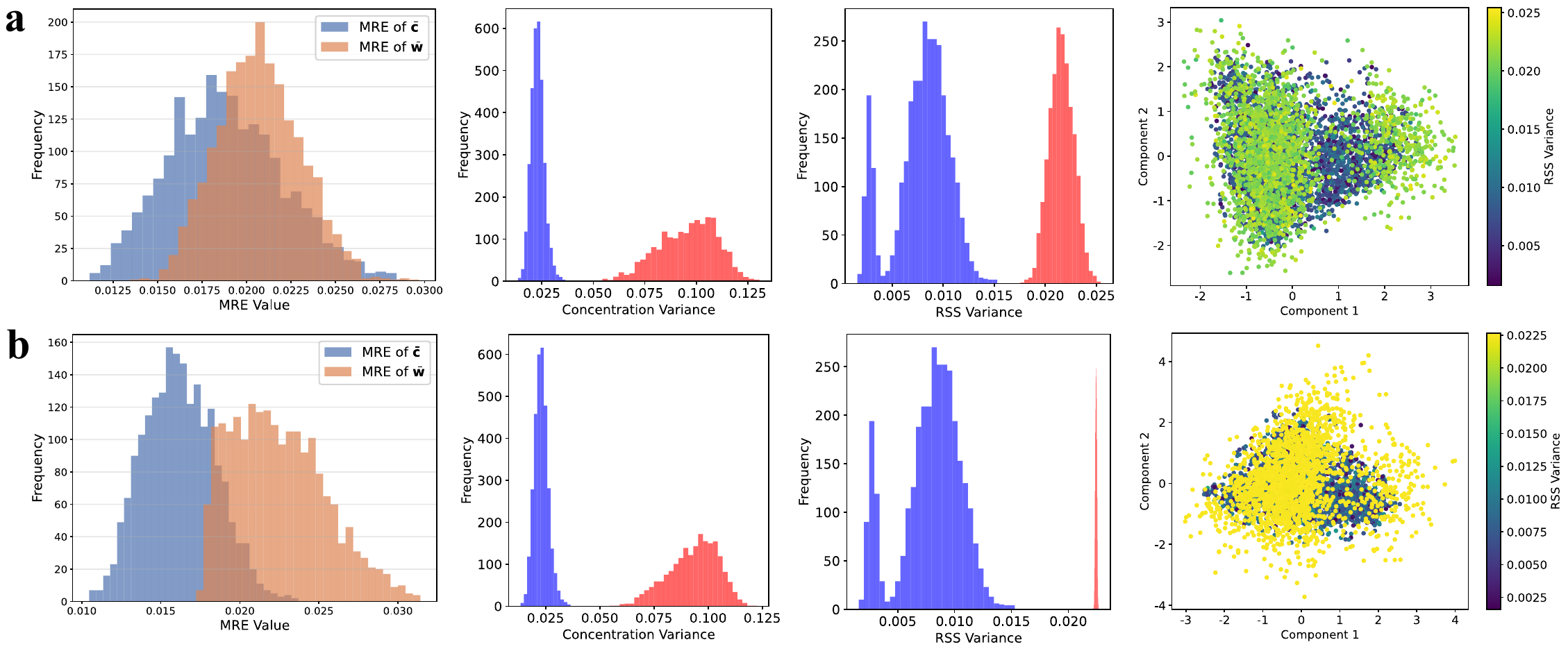}  
    \caption{Overall statistical information of two optimization scenarios. a) the Random Latent Method, and b) the Fixed Latent Method. The first column shows the histogram of self-checking MREs for 2000 samples at the last iteration step. All the MREs are less than 4\%. The second and third column give the histogram of the variance of concentration and variance of resolved shear stress for both training dataset (blue) and optimization results (red), respectively. The final column shows the Isomap results for the 2 optimization situations}
    \label{op_whole}
\end{figure}

\begin{figure}[htbp]
    \centering
    \includegraphics[scale=0.6]{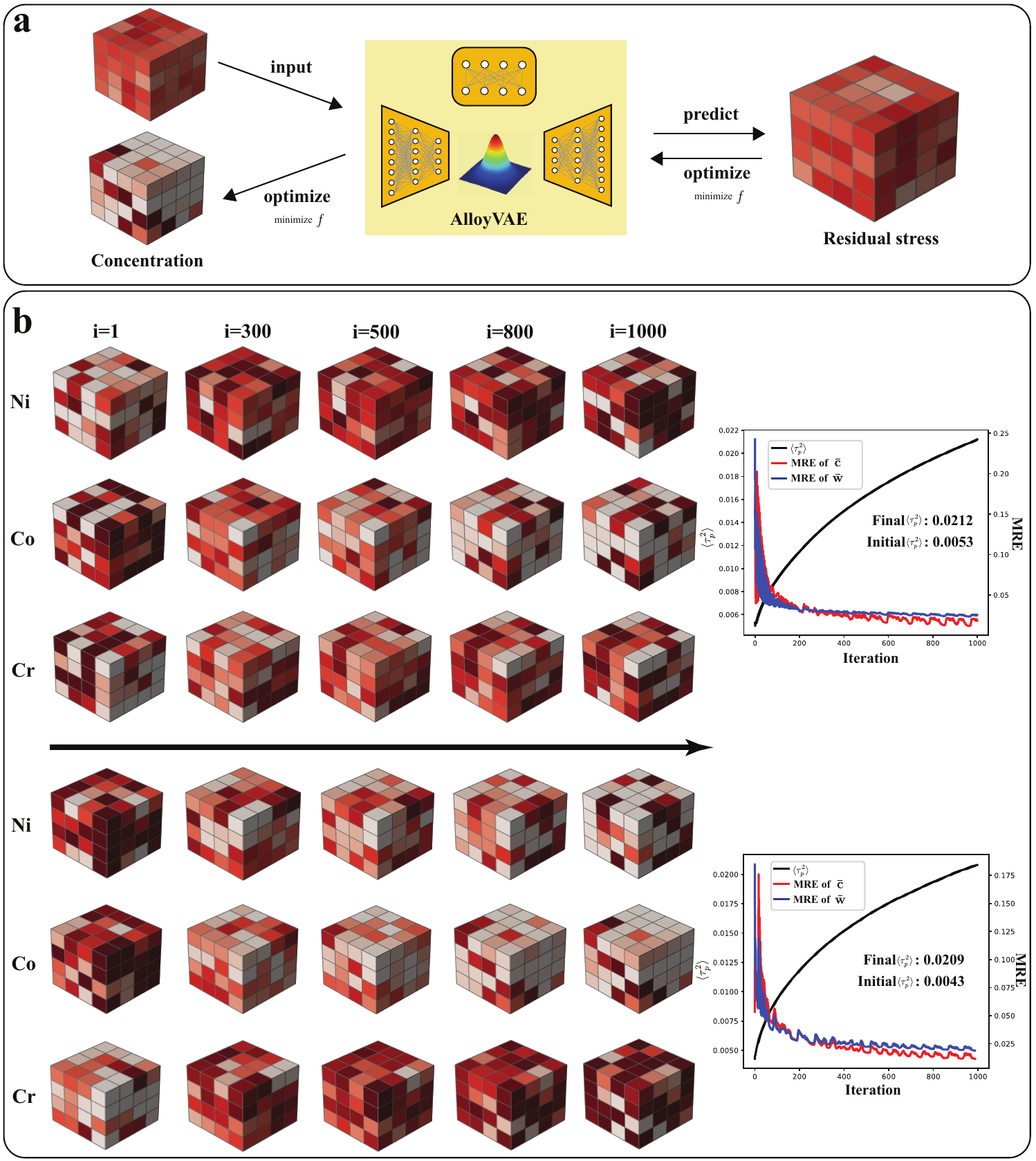}  
    \caption{The optimization process of concentration. (a) Schematic overview of the optimization algorithm based on the AlloyVAE framework, which receives concentration as input and obtains the optimized concentration by minimizing the objective function $f$. (b) The results of optimization process. The Ni, Co, Cr concentration distributions at iteration step i=1,300,500,800,1000 are given. The concentration of Cr is computed as $c_{\text Cr} = 1-(c_{\text Ni} + c_{\text Co})$), ensuring the total composition sums to unity. The  evolution of $\langle\tau_p^2\rangle$ (black solid line) and self-checking MRE (red and blue solid lines) are also illustrated in the right column.}
    \label{op_concen}
\end{figure}

We first evaluate all 2000 optimization results for the both latent variable initialization methods. As shown in Figure~\ref{op_whole}, the final self-checking MREs for all samples fall below 4\%, satisfying the accuracy requirement for valid stress prediction. The plots in the middle columns of Figure~\ref{op_whole} show a significant increase in the values of $\langle \tau_p^2 \rangle$. In particular, the optimization process initialized with random latent variables exhibits greater diversity in the results. Furthermore, the variance of optimized concentrations exceeds that of the original inputs, suggesting a useful concentration design strategy: a higher concentration variance leads to a greater variance in resolved shear stress. 

In addition, to visualize the structure of the optimized concentrations, we employed Isomap to project them onto a two-dimensional space. As shown in the last column of Figure~\ref{op_whole}, the resulting distribution reveals two prominent clusters in both scenarios. A closer analysis reveals that one cluster corresponds to a slight increase in the concentrations of Ni and Co, accompanied by a minor reduction in Cr from the initial equiatomic value of 0.5, while the other cluster exhibits the reverse trend. These subtle variations in elemental concentrations still satisfy the equiatomic condition, yet indicate a coupling effect between Ni and Co in contributing to the strengthening behavior.

Since the Random Latent Method generates more representative results, Figure~\ref{op_concen} further presents two examples of the concentration optimization process with different initializations of the latent variable (i.e., the Random Latent Method). The concentration of Cr is derived from the concentrations of Ni and Co, which are treated as independent variables in the optimization, by applying the compositional constraint that the sum of all three elemental concentrations equals unity. As the optimization progresses, the concentration distributions exhibit some fluctuations, while the value of $\langle \tau_p^2 \rangle$ increases steadily. During the process, the self-checking MREs for the concentration and SRO both decrease below 5\%, indicating that the resulting optimized concentrations are plausible and physically reasonable residual stress fields with enhanced $\langle \tau_p^2 \rangle$ are generated.

\subsection{Application with eigenstrain}
\label{sec:eigenstrain}
The residual stress prediction and inverse concentration design presented above represent a primary application of the generalized AlloyVAE framework, which is design to learn the mapping from structure properties $\mathbf{p}_1, \mathbf{p}_2, ..., \mathbf{p}_N$ to mechanical properties $\mathbf{m}_1, \mathbf{m}_2, ..., \mathbf{m}_M$. This flexible formulation allows multiple structural and mechanical variables to be simultaneously input to the encoder and smoother, respectively, with the total training loss adapted to the chosen variables. By adjusting the input and output configurations and corresponding loss terms, the model can be readily extended to other complex problems in materials science.


To demonstrate the ability of AlloyVAE to migrate to other MPEA problems, we introduce the concept of eigenstrain and train another two models to bridge composition fields, eigenstrain and residual stress. Different compositions fields will lead to different equilibrium volumes, and the volume mismatch will form an eigenstrain field, which will finally influence the residual stress. Moreover, eigenstrain offers a carrier to future model generalization to rigorously incorporate defects in MPEAs, thereby enabling us to model complex, realistic scenarios beyond our current defect-free systems. We make minor adjustments to the framework in Fig.\ref{general_framework} (a), and give the other two new frameworks in the supplementary to solve different tasks. The first framework establishes a surrogate mapping from composition fields to eigenstrain, while the second framework models the combined influence of composition fields and eigenstrain on residual stress. Both of these continuum-scale relationships are physically meaningful, but there is currently no effective mathematical model to accurately characterize them. These two frameworks are modified at input \& output variables of different network modules and model architectures slightly, and they are trained with datasets containing eigenstrain calculated from MD simulations (see details in the Methods Section 4.1.2).

\begin{figure}[htbp]
    \centering
    \includegraphics[scale=0.3]{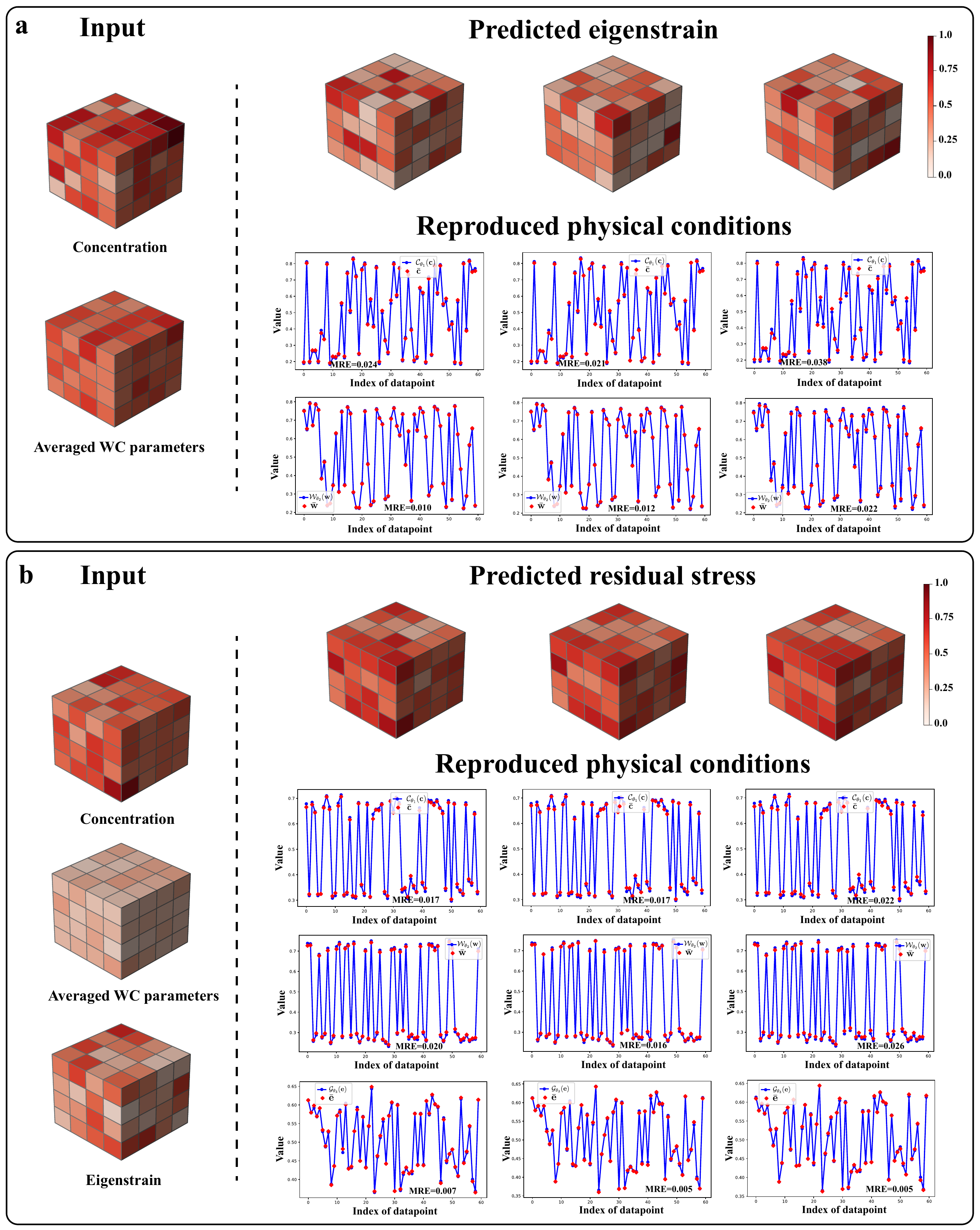}  
    \caption{Predictions for eigenstrain and residual stress through a self-consistent process. a) Eigenstrain (\(xx-\)component) predictions from concentration and averaged WC parameters. 
    b) Residual stress (\(xx-\)component) predictions from concentration, averaged WC parameters, and eigenstrain. Each subfigure shows the inputs on the left, the predictions in the upper right, and the reproduced physical conditions in the lower right. Both frameworks reproduce physical conditions with mean relative errors below 4\%. The dimension of physical condition is 40.}
    \label{2framework results}
\end{figure}

The performance of validity of these two frameworks are given in supplementary, indicting all MREs of reconstruction and physical conditions predictions are around or less than 5\% in test dataset, which guarantees the effectiveness. Fig. \ref{2framework results} shows the performance of eigenstrain and residual stress predictions from a specific composition field or eigenstrain inputs of these two frameworks using the self-checking process. All the predictions can reproduce the physical conditions with MRE less than the threshold 4\%, providing meaningful references in practice. These results demonstrate the transferability of AlloyVAE across different datasets, highlighting its adaptability to related materials science problems.

\section{Discussion}
In this work, we have introduced AlloyVAE as a generative, physics-informed framework for modeling structure–property relationships in multi-principal element alloys. By learning the conditional distribution of mechanical fields from coarse-grained microstructural descriptors, the framework addresses a fundamental limitation of existing approaches: the implicit assumption that structure uniquely determines property. Our results demonstrate that, once atomistic information is projected onto composition and short-range order fields, the resulting mapping to mechanical response is inherently non-unique. AlloyVAE resolves this challenge by explicitly representing such relationships in a probabilistic manner, enabling the generation of multiple physically admissible realizations under identical input conditions.

The proposed framework integrates several key elements that collectively enable this capability. First, the conditional variational autoencoder provides a natural representation of one-to-many mappings through structured latent variables, allowing efficient sampling of diverse mechanical field configurations. Second, the introduction of learned smoothers enhances the regularity of the underlying functional mapping, effectively acting as a data-driven coarse-graining operator that stabilizes training and improves generalization. Third, the self-consistency mechanism through self-checking enforces physical plausibility by rejecting predictions that are incompatible with the prescribed input conditions. Together, these components establish a robust and scalable approach for probabilistic field-to-field modeling in heterogeneous materials.

Beyond forward prediction, AlloyVAE enables inverse design by navigating the high-dimensional compositional space through gradient-based optimization. By coupling the generative model with physically motivated objective functions, we demonstrate the ability to identify composition fields that enhance dislocation resistance via increased stress fluctuations. This capability highlights the potential of probabilistic generative models not only as surrogates for expensive simulations, but also as design engines for discovering non-intuitive microstructural configurations. Importantly, the diversity of generated solutions reflects the underlying multiplicity of physically admissible states, offering a richer design landscape than deterministic approaches.

The extension of the framework to include eigenstrain further underscores its generality and connection to classical micromechanics. Eigenstrain serves as a unifying descriptor linking composition, lattice distortion, defects, and residual stress. This incorporation enables AlloyVAE to bridge multiple levels of physical descriptions. This opens the door to modeling more realistic scenarios involving defects, phase transformations, and thermo-mechanical histories, where traditional analytical or numerical approaches remain challenging. More broadly, the ability to learn coupled field relationships suggests that the framework can be extended to other spatially distributed properties, including elastic moduli, heat flux, and charge density.

From a broader perspective, this work points toward a shift in how structure–property relationships are conceptualized in complex materials. In heterogeneous systems subject to coarse-graining, variability is not merely noise but an intrinsic feature arising from unresolved degrees of freedom. Deterministic models, whether physics-based or data-driven, necessarily collapse this variability and may therefore overlook critical aspects of material behavior, particularly those associated with extreme events such as failure initiation. By contrast, probabilistic representations capture the full distribution of possible responses, enabling the analysis of variability, uncertainty, and rare but consequential outcomes.

The AlloyVAE framework thus provides a foundation for a probabilistic mechanics paradigm, in which material behavior is described not by a single deterministic response but by a distribution conditioned on microstructural state. Such a perspective is particularly relevant for the design of advanced materials with engineered heterogeneity, where performance may depend sensitively on fluctuations and local extremes. By combining high-fidelity simulation data with scalable generative modeling, the approach offers a practical pathway for exploring high-dimensional design spaces that are otherwise computationally inaccessible.

Looking forward, several directions may further enhance the scope and impact of this framework. Integrating experimental data, incorporating explicit physical constraints such as equilibrium or compatibility conditions, and extending the latent representations to account for temporal evolution under loading are all promising avenues. In addition, coupling the model with uncertainty quantification and reliability analysis could enable predictive assessment of failure probabilities in complex materials systems. These developments would further strengthen the role of probabilistic generative modeling as a core tool in materials science. 

In summary, by formulating structure-property relationships as probabilistic field mappings and demonstrating their practical realization through AlloyVAE, this work establishes a new framework for modeling and designing heterogeneous materials. This paradigm complements and extends traditional deterministic approaches, offering new opportunities to understand, predict, and engineer material behavior in systems where variability is fundamental rather than incidental.

\section{Methods}
\subsection{Generation of Atomistic and Coarse-Grained Data}
\subsubsection{Stress dataset}
We present the setup of the atomistic simulations. Equiatomic CoCrNi simulation cells are constructed with dimensions of 302 nm in the $x$-direction, 305 nm in the $y$-direction, and 308 nm in the $z$-direction, containing a total of 2,520,000 atoms. The simulations are performed to generate 3500 different atomistic samples using the open-source Large-scale Atomic/Molecular Massively Parallel Simulator (LAMMPS)~\cite{LAMMPS}, with the embedded atom method (EAM) interatomic potential for the Co-Cr-Ni system developed by Li et al.\cite{li2019nc}. 

Random simulation samples were directly generated in LAMMPS with different random seeds. To obtain samples with short-range order (SRO), a hybrid MD/Monte Carlo (MC) approach was employed at annealing temperatures ranging from 600 K to 1100 K \cite{jian2020acta,xie2021acta}. During this process, the MD timestep was set to 2.5 fs within the NPT ensemble, with periodic boundary conditions (PBC) applied in all three directions while maintaining the system at the prescribed annealing temperature \cite{li2019nc}. An MC cycle was conducted every 20 MD steps to attain a more stable configuration, during which 25\% of the atoms were selected for potential swaps based on the Metropolis algorithm to minimize the system energy. This procedure continued until equilibrium configurations with the target compositions and thermodynamically stable atomic occupations were achieved. The resulting SRO-induced samples were then quenched to 1 K, followed by further energy minimization until all three normal stress components reached zero. To quantify the degree of SRO, the 1st order WC parameters for pairwise atomic interactions are calculated as  
$\alpha_{i j}= \frac{p_{i j}-c_{j}}{\delta_{i j}-c_{j}}$. Here $p_{i j}$ denotes the probability of a $j$-type atom being around an atom of type $i$ within the 1st nearest-neighbor shell, $c_{j}$ is the concentration of $j$-type atom, and $\delta_{i j}$ is the Kronecker delta function \cite{cowley1950pr,de197jac}.

To generate a coarse-grained dataset beyond the atomistic data for practical applications, the residual stress dataset generated from the atomistic simulations is processed to obtain spatially averaged concentrations, SRO levels, and stress components. A block-averaging approach is applied to convert the dataset to a continuum scale by partitioning the simulation domain into a $4 \times 4 \times 4$ grid along each spatial direction. Thus, each grid cell is nearly cubic, with an approximate edge length of 75 nm. For each grid, the concentration of each element is computed as the fraction of atoms of that element relative to the total number of atoms in this grid. After computing the concentration for all grids, we can get the concentration field $\mathbf{c}_i$ for the sample $i$. The SRO level of sample $i$ is determined by averaging the Warren-Cowley (WC) parameters 
within each grid and the averaged WC parameters $\mathbf{w}_i$ is obtained. Similarly, the residual stress field of $i$ th sample $\boldsymbol{\sigma}_i$ is approximated by spatially averaging the Virial stress over each grid. The obtained coarse-grained dataset for ML models input is represented as a high-dimensional tensor, where the concentration $\mathbf{c}$, the average WC parameter $\mathbf{w}$, and the residual stress $\boldsymbol{\sigma}$ have dimensions of $(b_n, 2, 4, 4, 4)$ (only 2 independent concentration variables), $(b_n, 9, 4, 4, 4)$, and $(b_n, 6, 4, 4, 4)$, respectively, with $b_n$ denoting the batch size. To ensure consistency and comparability, all these concentration $\mathbf{c}$, average WC parameters $\mathbf{w}$ and residual stress $\boldsymbol{\sigma}$ are normalized to the $[0,1]$ range using min-max normalization. Finally, the dataset is divided into 90\% for training and 10\% for testing. 

\subsubsection{Eigenstrain dataset}
Similarly, the eigenstrain dataset is derived from atomistic simulations using the same atomic samples and grid cells employed in the stress data generation. The embedded volume of each grid cell within the sample can be directly computed, whereas the corresponding stress-free volumes remain unknown. Therefore, an additional set of atomistic simulations is performed, where each grid cell was sequentially extracted from the sample and fully relaxed via energy minimization under free boundary conditions. The eigenstrain is then computed as the deviation from unity of the cube root of the ratio between the relaxed, stress-free volume and the original embedded volume. This process yields both stress and eigenstrain data for a consistent set of grid cell partitions.

\subsection{VAE based framework}
Traditional VAE \cite{kingma2013auto, doersch2016tutorial,kingma2015variational} is a powerful generative model that has been successfully used in numerous fields such as images, videos, and speech. Traditional VAE is composed of an encoder and decoder, both of which are essentially neural networks. Neural networks are computational models inspired by the human brain, consisting of interconnected layers of nodes (or neurons) that process and transmit information. They are widely used in machine learning for tasks such as pattern recognition, classification, and prediction \cite{xiong2022physically, du2021two, zhao2021hot, santhosh2021vehicular,rampavsek2019dr}. By adjusting the weights of connections through training, neural networks can learn to approximate complex functions and make data-driven decisions. VAE uses the encoder to encode the input into a latent space, and a latent variable $\mathbf{z}$, which is randomly sampled from this latent space (reparameterization trick), is fed into the decoder to reconstruct the input. 

In the generate phase, we only need to randomly sample the latent variables $\mathbf{z}$ and input them to the decoder to generate results. VAE has an elegant probabilistic nature \cite{kingma2013auto}, which guarantees a better diversity of outputs and a relatively low computational cost compared to other generative models such as generative adversarial networks (GAN) \cite{goodfellow2020generative}. However, traditional VAE lacks control over the generation process, thus the conditional VAE (cVAE) \cite{zhao2017learning,kim2021conditional} is proposed to give both the encoder and decoder another input $\mathbf{y}$. The cVAE can efficiently generate results under the guidance of specific conditions, which is more in line with the setting of our problem. Therefore, we design an ML framework similar to cVAE as shown in Figure\ref{VAE framework}. The goal of the framework is to maximize the log-likelihood of the conditional probability under physical conditions $\mathbf{y}=(\bar{\mathbf{c}},\bar{\mathbf{w}})$
\begin{align}
    \mathrm{log}\mathcal{D}_q (\boldsymbol{\sigma}|\mathbf{y})= &\mathbb{E}_{\mathcal{E}_\phi(\mathbf{z}|\boldsymbol{\sigma},\mathbf{y})}[\mathrm{log}\mathcal{D}_q(\boldsymbol{\sigma}|\mathbf{y})]\nonumber=\mathbb{E}_{\mathcal{E}_\phi(\mathbf{z}|\boldsymbol{\sigma},\mathbf{y})}[\mathrm{log}\frac{\mathcal{D}_q(\boldsymbol{\sigma},\mathbf{z}|\mathbf{y})}{\mathcal{D}_q(\mathbf{z}|\boldsymbol{\sigma},\mathbf{y})}]\nonumber\\
    =&\mathrm{ELBO}+D_{KL}(\mathcal{E}_\phi(\mathbf{z}|\boldsymbol{\sigma},\mathbf{y})\|\mathcal{D}_q(\mathbf{z}|\boldsymbol{\sigma},\mathbf{y}))\nonumber\\
    \geqslant&\mathrm{ELBO}\quad(\mathrm{since}\ 
 D_{KL}\geqslant0)\nonumber\\
=&\mathbb{E}_{\mathcal{E}_\phi(\mathbf{z}|\boldsymbol{\sigma},\mathbf{y})}[\mathrm{log}\mathcal{D}_q(\boldsymbol{\sigma}|\mathbf{z},\mathbf{y})]-D_{KL}(\mathcal{E}_\phi(\mathbf{z}|\boldsymbol{\sigma},\mathbf{y})\|\mathcal{E}(\mathbf{z}|\mathbf{y}))
    \label{VAE def}
\end{align}
where we assume the prior $\mathcal{E}(\mathbf{z}|\mathbf{y})\sim\mathcal{N}(\mathbf{0},\mathbf{I})$ and $\mathcal{D}_q(\boldsymbol{\sigma}|\mathbf{z},\mathbf{y})$ is also a normal distribution. We can make the framework learn the distribution of residual stress under specific conditions by maximizing the ELBO, and the last term of (\ref{VAE def}) 
is related to the first two terms (reconstruction loss and KL divergence) of total loss (\ref{loss}) using the Gaussian assumptions above.  The detailed architectures of the encoder, decoder and two smoothers are illustrated in supplementary material, and the total number of parameters of the framework is around 2.8$\times10^8$. The entire ML framework is implemented in PyTorch\cite{paszke2019pytorch} with Xavier parameters initialization, and the Adam optimizer\cite{kingma2014adam} is adopted to optimize the loss function (\ref{loss}). The framework is trained on a single NVIDIA GeForce RTX 4090 with 24GB memory in an end-to-end manner, along with a batch size of 200 and 1000 training epochs.

\section*{Acknowledgments}
The work of Y.X. was supported by National Key Research and Development Program of China (No. 2025YFA1016800).
\bibliographystyle{unsrt}
\bibliography{references}

\newpage
\begin{center}
    \Large\bfseries Supplementary Information
\end{center}
\setcounter{section}{0} 
\renewcommand{\thesection}{\Alph{section}}

\section{Block-average process}
\begin{figure}[htbp]
	\centering
	\includegraphics[scale=0.45]{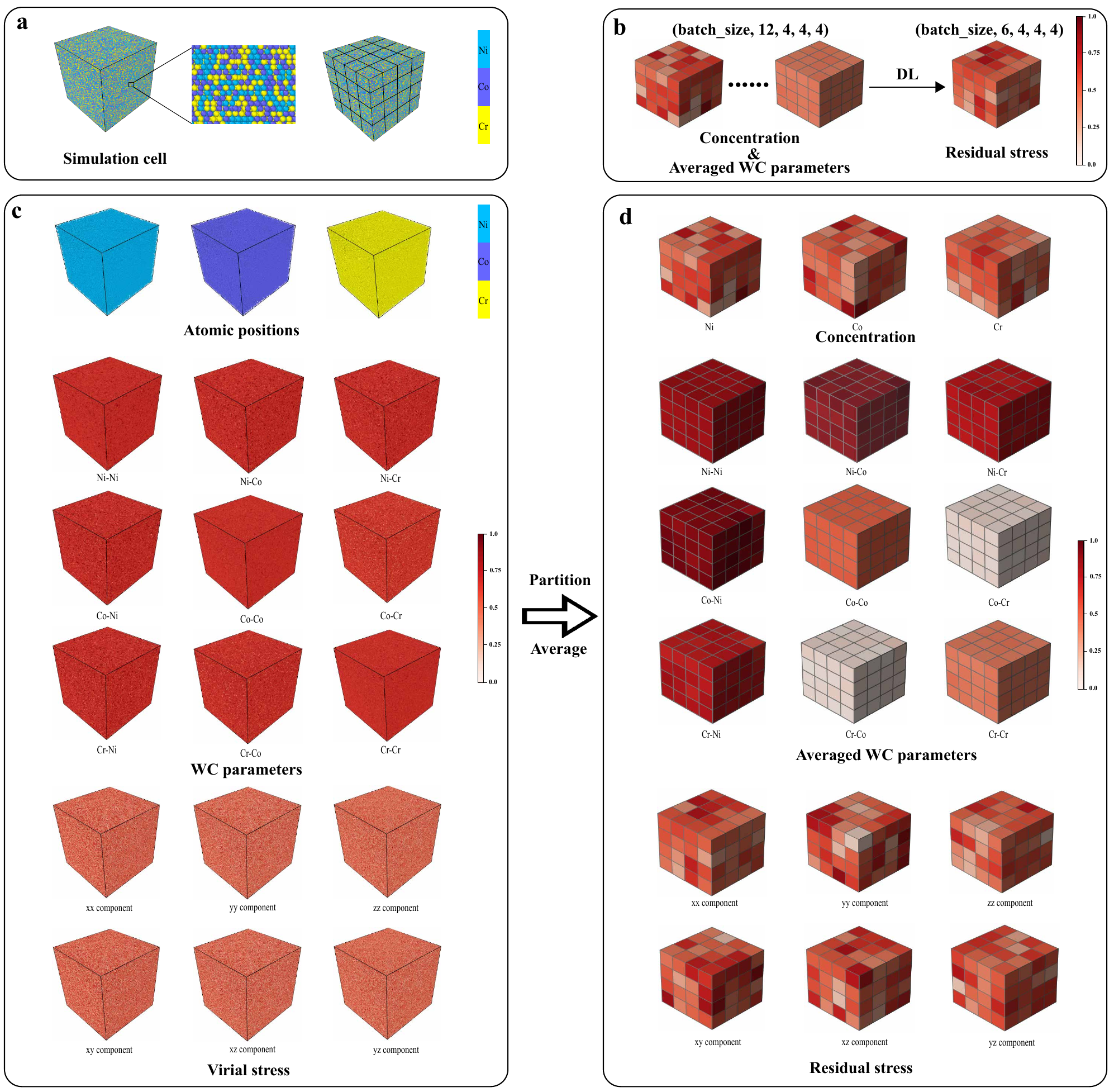}
	\caption{Overview of the dataset. a) The partition of the simulation cell using (4,4,4) resolution. b) The input concentration and averaged WC parameters as well as output residual stress tensor size of DL model. The concentration tensor size is ($b_n$,2,4,4,4), averaged WC parameters size is ($b_n$,9,4,4,4) and residual stress size is ($b_n$,6,4,4,4). c) MD simulation results. d) The dataset obtained by partition and average on MD results.}
	\label{atomic to continuum}
\end{figure}

Figure \ref{atomic to continuum} illustrates the whole process of block-average. The key idea is partitioning each dimension of the simulation cell into many small blocks, where each block contains numerous atoms. Then the mechanical fields, including atomic arrangement of 3 elements, WC parameters of 9 atom pairs and virial stress of 6 directions, are averaged in each block to get the representative volume element (RVE), then we will get the concentration, averaged WC parameters and residual stress. This partition and average process can be considered as a data transformation from microscale to continuum scale. The DL tensor size of concentration is (${b_n}\times2\times4\times4\times4$), considering only 2 independent concentration variables, the size of averaged WC parameter is (${b_n}\times9\times4\times4\times4$) and stress tensor is (${b_n}\times6\times4\times4\times4$), where ${b_n}$ is the batch size for neural work training. We aim to predict residual stress from element concentration and averaged WC parameters using DL models.

It is worth noting that this data generation method based on partition-average will influence the generalization ability of DL models. Roughly speaking, this process will lose the high-frequency details for both input and output, making it difficult for DL models to capture detailed patterns of input\cite{koziarski2018impact,wang2016studying,al2023effects}. Intuitively, the input samples may present the trend of clustering and will not completely fill the sample space. One potential result is a large vibration of output resulting from a small perturbation on input, which will increase the model output variance and further increase the generalization error of DL models on test dataset. In this case, traditional supervised learning models with MSE loss will have bad performance on test dataset. So we need to choose the appropriate ML model to complete this regression problem.

\section{One-to-many mapping}
\begin{figure}[htbp]
	\centering
    \includegraphics[scale=.4]{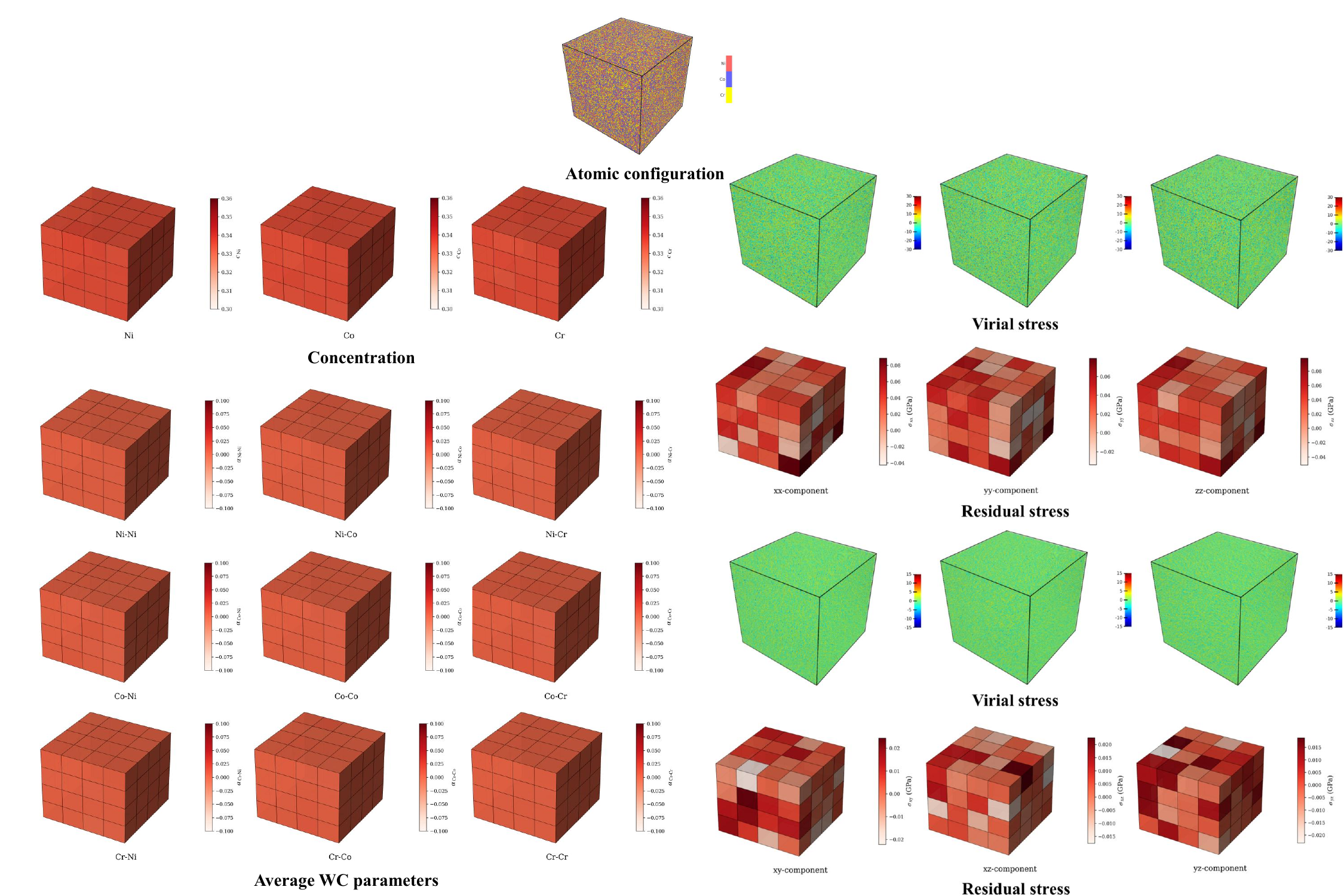}
	\caption{Atomic positions, concentrations, average WC parameters, corresponding virial stresses and residual stresses of concentration of MD Sample 1.}
	\label{one2many_1}
\end{figure}

\begin{figure}[htbp]
	\centering
    \includegraphics[scale=.4]{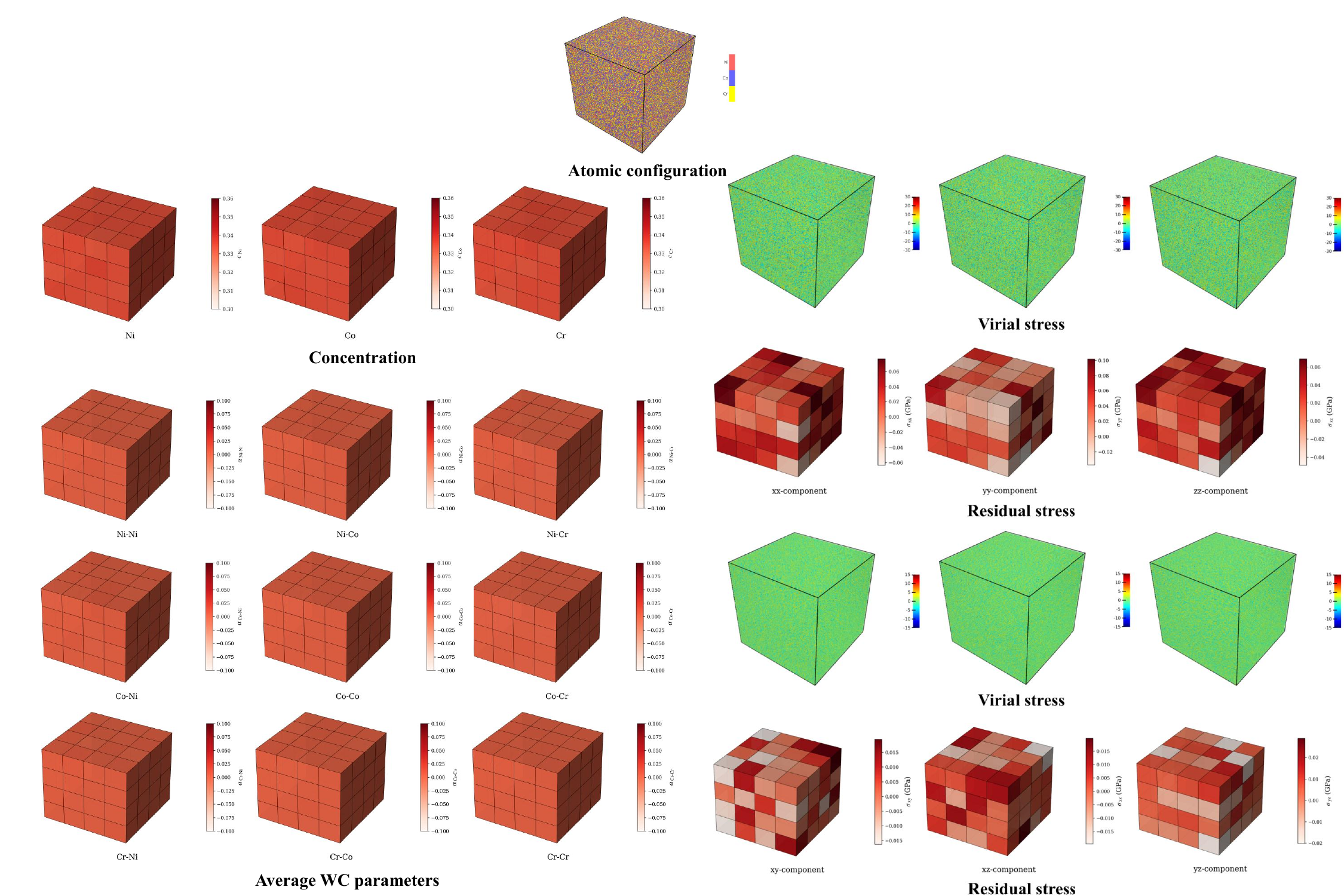}
	\caption{Atomic positions, concentrations, average WC parameters, corresponding virial stresses and residual stresses of concentration of MD Sample 2.}
	\label{one2many_2}
\end{figure}

\begin{figure}[htbp]
	\centering
    \includegraphics[scale=.4]{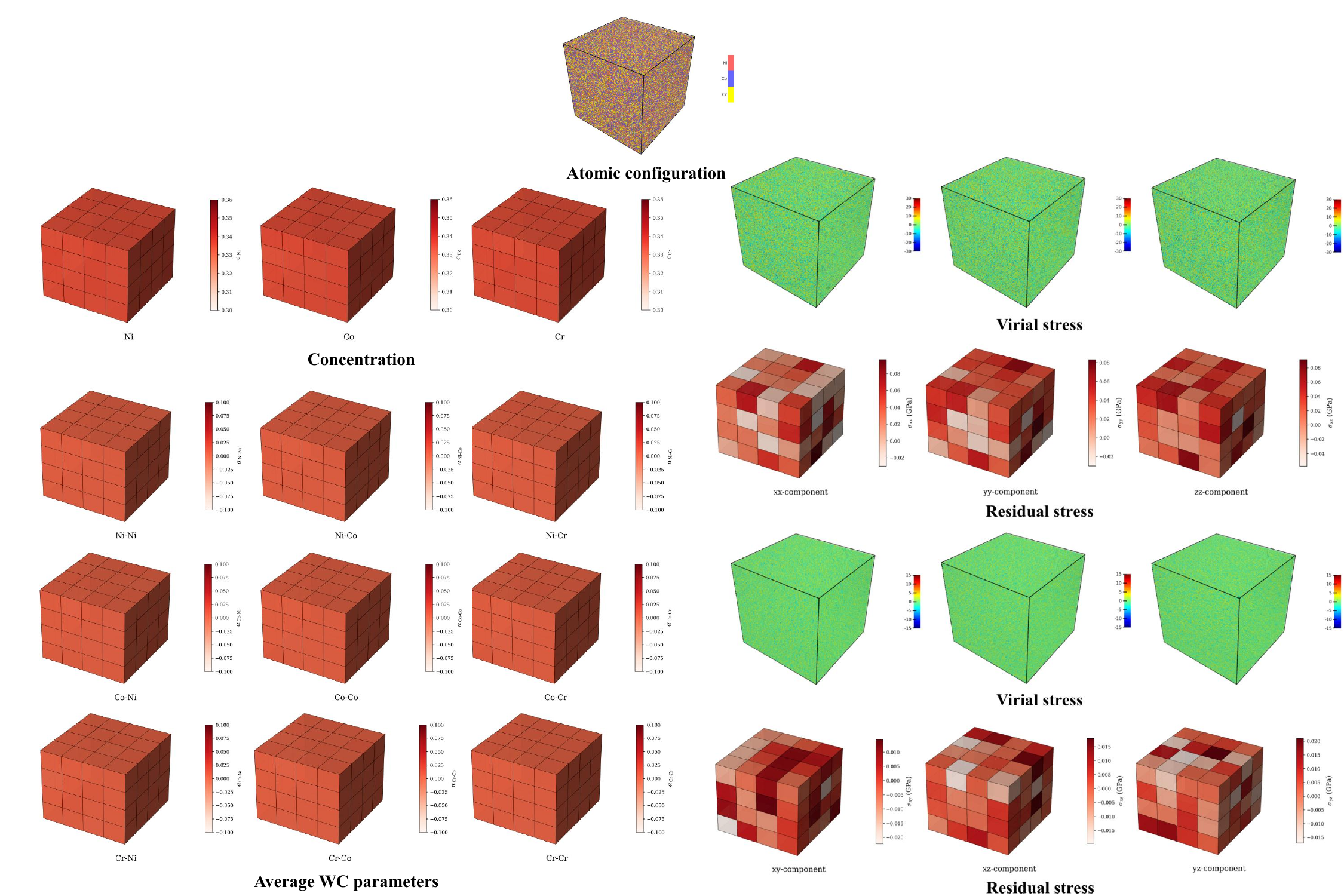}
	\caption{Atomic positions, concentrations, average WC parameters, corresponding virial stresses and residual stresses of concentration of MD Sample 3.}
	\label{one2many_3}
\end{figure}

To demonstrate the one-to-many mapping caused by spatial averaging and the subsequent loss of atomic configurational detail, three random MD samples were generated with equi-atomic concentrations and near-zero averaged WC parameters (within numerical tolerance). As shown in Figs.~\ref{one2many_1}–\ref{one2many_3}, despite the identical continuum descriptors of the three MD samples, their residual stress fields differ significantly, even exhibiting reversals between compression and tension at certain regions.

\section{Neural Network Architecture}
Table \ref{tab:network_architecture} lists the detailed neural network architecture of encoder, two smoother and decoder for the main framework Fig. 2 in the main article. As for the encoder, 3 layers 3D convolution followed by a fully connected (Fc\_feature) layer are used to transfer the residual stress tensor into a feature vector with the size of 4×4×4×64. Then 2 FNN sequences use this feature vector to predict physical condition $\bar{\mathbf{c}},\bar{\mathbf{w}}$, respectively. $\bar{\mathbf{c}},\bar{\mathbf{w}}$, together with this feature vector are then transferred into the mean value vector $\boldsymbol{\mu}$ and diagonal covariance vector $\boldsymbol{\Sigma}$ with 2 FNN sequences. The architectures of two smoothers are similar, starting with a 3D convolution layer followed by 2 fully connected layers, which finally predict the physical conditions. The decoder is composed of 5 fully connected layers. The latent space size is set to 300, while the sizes of the physical conditions $\bar{\mathbf{c}}$ and $\bar{\mathbf{w}}$ are both set to 150. We also add batch normalization and sigmoid activation layer after neural network layers.

\begin{table}[htbp]
\centering

\begin{tabular}{|c|c|c|c|c|}
\hline
\textbf{network} & \textbf{layer} & \textbf{size-in} & \textbf{size-out} & \textbf{kernel} \\
\hline
 & Input stress& - & 4×4×4×6 & - \\
\hline
\multirow{15}{*}{Encoder} & Conv\_1 & 4×4×4×6 & 4×4×4×32 & 3×3×3×32 \\
\cline{2-5}
 & Conv\_2 & 4×4×4×32 & 4×4×4×64 & 3×3×3×64 \\
\cline{2-5}
 & Conv\_3 & 4×4×4×64 & 4×4×4×128 & 3×3×3×128 \\
\cline{2-5}
 & Fc\_feature & 4×4×4×128 & 4×4×4×64 & - \\
\cline{2-5}
 & Fc\_concen1 & 4×4×4×64 & 4×4×4×16 & - \\
\cline{2-5}
 & Fc\_concen2 & 4×4×4×16 & 4×4×4×8 & - \\
\cline{2-5}
 & Fc\_concen3 & 4×4×4×8 & 150 & - \\
\cline{2-5}
 & Fc\_wc1 & 4×4×4×64 & 4×4×4×16 & - \\
\cline{2-5}
 & Fc\_wc2 & 4×4×4×16 & 4×4×4×8 & - \\
\cline{2-5}
 & Fc\_wc3 & 4×4×4×8 & 150 & - \\
 \cline{2-5}
 & Fc\_mu1 & 4×4×4×64+150×2 & 4×4×4×16 & - \\
  \cline{2-5}
 & Fc\_mu2 & 4×4×4×16 & 300 & - \\
  \cline{2-5}
 & Fc\_logvar1 & 4×4×4×64+150×2 & 4×4×4×16 & - \\
  \cline{2-5}
 & Fc\_logvar2 & 4×4×4×16 & 300 & - \\
\hline
\multirow{3}{*}{Smoother\_c} & Conv\_c1 & 4×4×4×2 & 4×4×4×64 & 3×3×3×64 \\
\cline{2-5}
 & Fc\_c1 & 4×4×4×64 & 4×4×4×32 & - \\
\cline{2-5}
 & Fc\_c2 & 4×4×4×32 & 150 & - \\
 \hline
\multirow{3}{*}{Smoother\_w} & Conv\_w1 & 4×4×4×9 & 4×4×4×64 & 3×3×3×64 \\
\cline{2-5}
 & Fc\_w1 & 4×4×4×64 & 4×4×4×32 & - \\
\cline{2-5}
 & Fc\_w2 & 4×4×4×32 & 150 & - \\
\hline
\multirow{4}{*}{Decoder} & Fc\_d1 & 300+150×2 & 4×4×4×128 & - \\
\cline{2-5}
 & Fc\_d2 & 4×4×4×128 & 4×4×4×128 & - \\
\cline{2-5}
 & Fc\_d3 & 4×4×4×128 & 4×4×4×128 & - \\
\cline{2-5}
 & Fc\_d4 & 4×4×4×128 & 4×4×4×128 & - \\
\cline{2-5}
 & Fc\_d5 & 4×4×4×128 & 4×4×4×6 & - \\
\hline
\end{tabular}
\caption{Summary of Network Architecture}
\label{tab:network_architecture}
\end{table}

\section{Additional results of AlloyVAE performance}
\begin{figure}[htbp]
	\centering
	\includegraphics[scale=0.103]{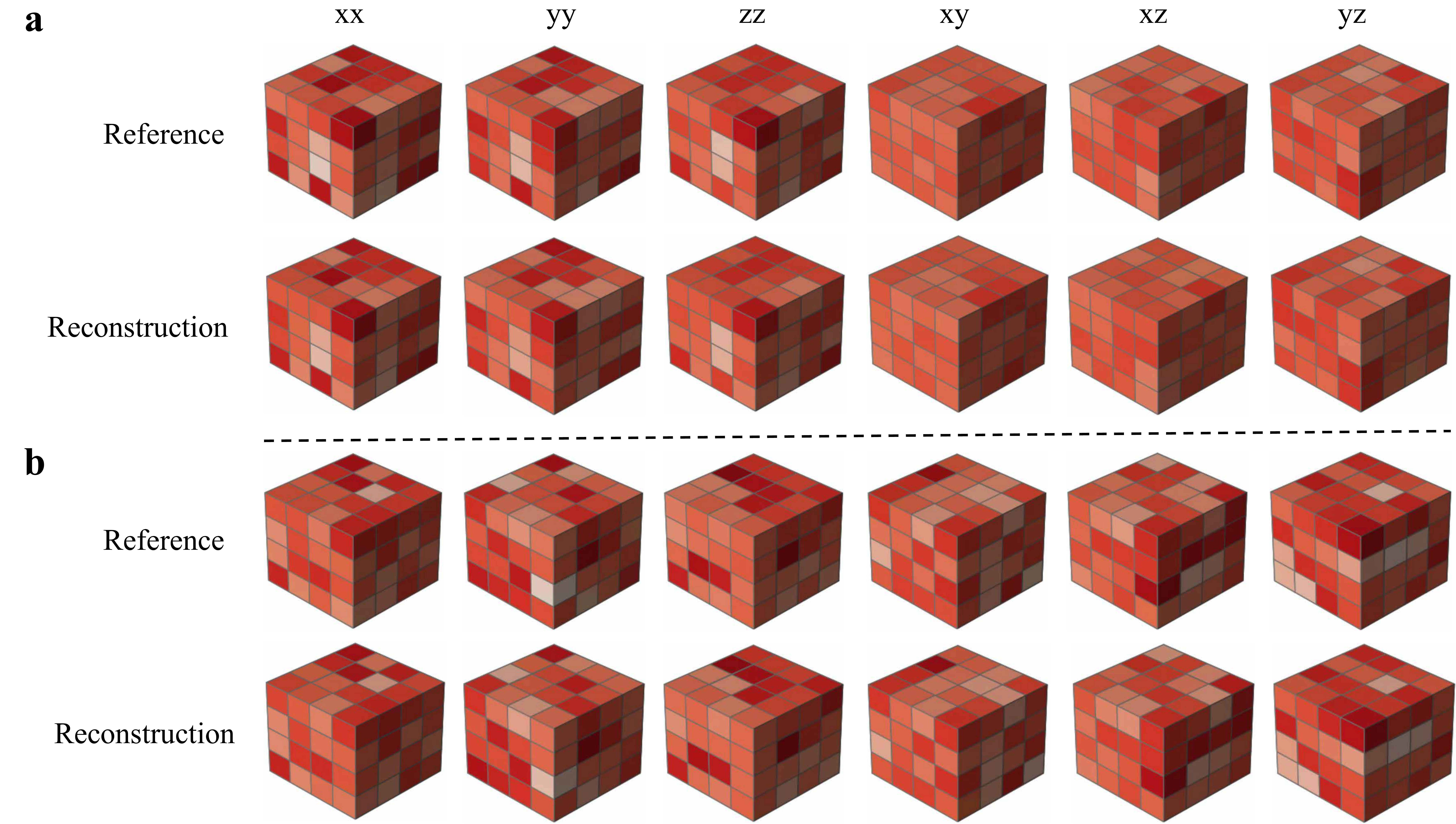}
	\caption{2 randomly chosen residual stress reconstruction examples with full 6 directions of stress.}
	\label{reconstruction}
\end{figure}

\begin{figure}[htbp]
	\centering
	\includegraphics[scale=0.45]{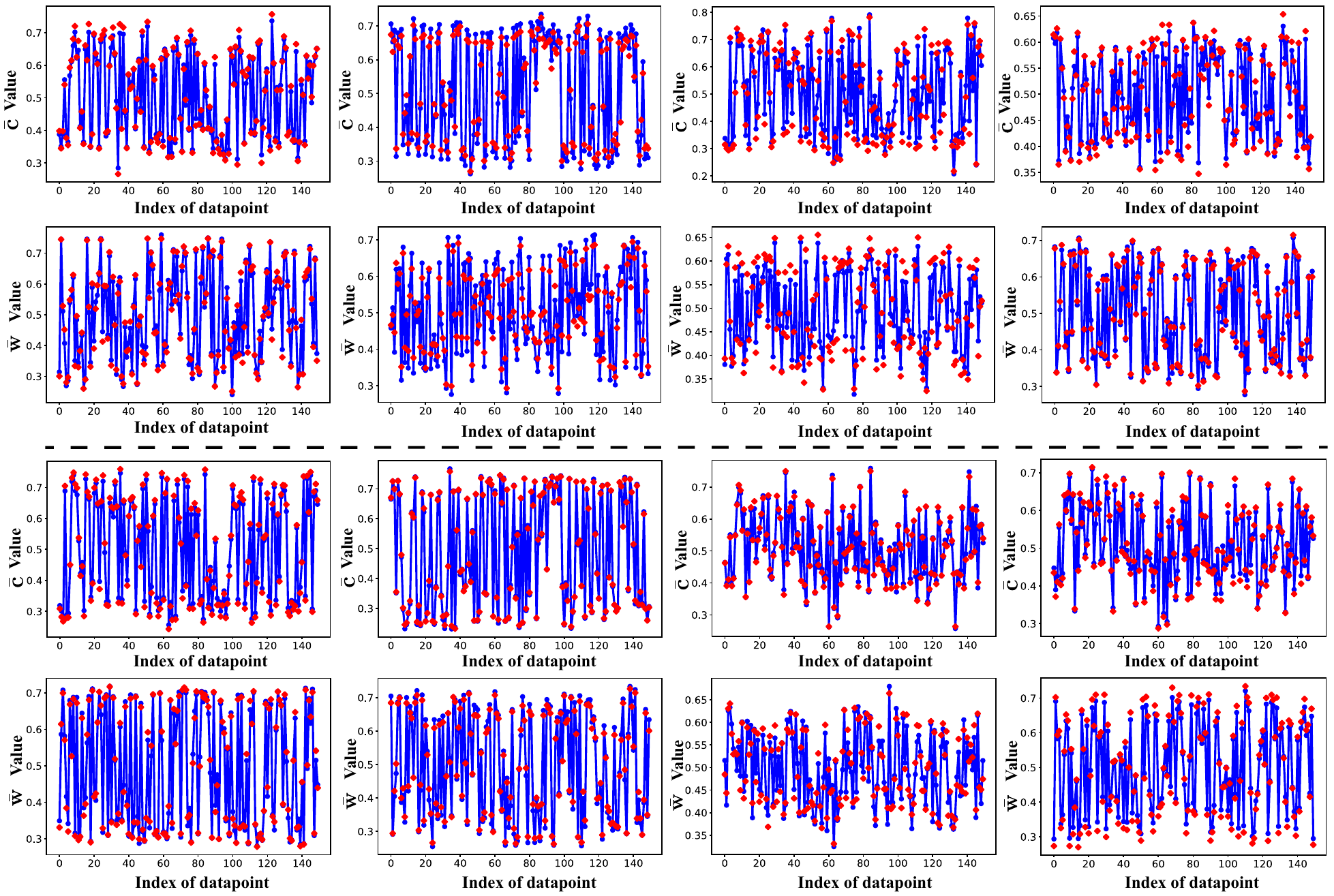}
	\caption{8 randomly chosen examples of physical predictions. Blue line is the reference while red dot is the encoder prediction}
	\label{physical prediction}
\end{figure}

We also focus on the additional results for the main demonstration Fig. 2 in the main article. In Figure \ref{reconstruction}, we illustrate additional 2 random examples of residual stress reconstruction in full 6 directions. This indicates that our framework can reconstruct the total stress information accurately and stably without disparity under some specific directions. Figure \ref{physical prediction} shows 8 randomly chosen predictions of physical condition, which match the reference very well with MRE $<$ 5\%. Different physical condition prediction shows specific fluctuation patterns instead of just horizontal line, indicating our framework overcomes the mode collapse and learns the intrinsic essence of concentration and averaged WC parameters.

\section{Additional results of residual stress prediction}
We can predict more scenarios of residual stress. Here we randomly choose 2 concentrations and averaged WC parameters pairs from test dataset, and predict 5 residual stress situations. All these 5 predictions can reconstruct the physical conditions with MRE below 4\%, so these predictions are reliable references for practical applications.
\begin{figure}[htbp]
	\centering
	\includegraphics[scale=0.4]{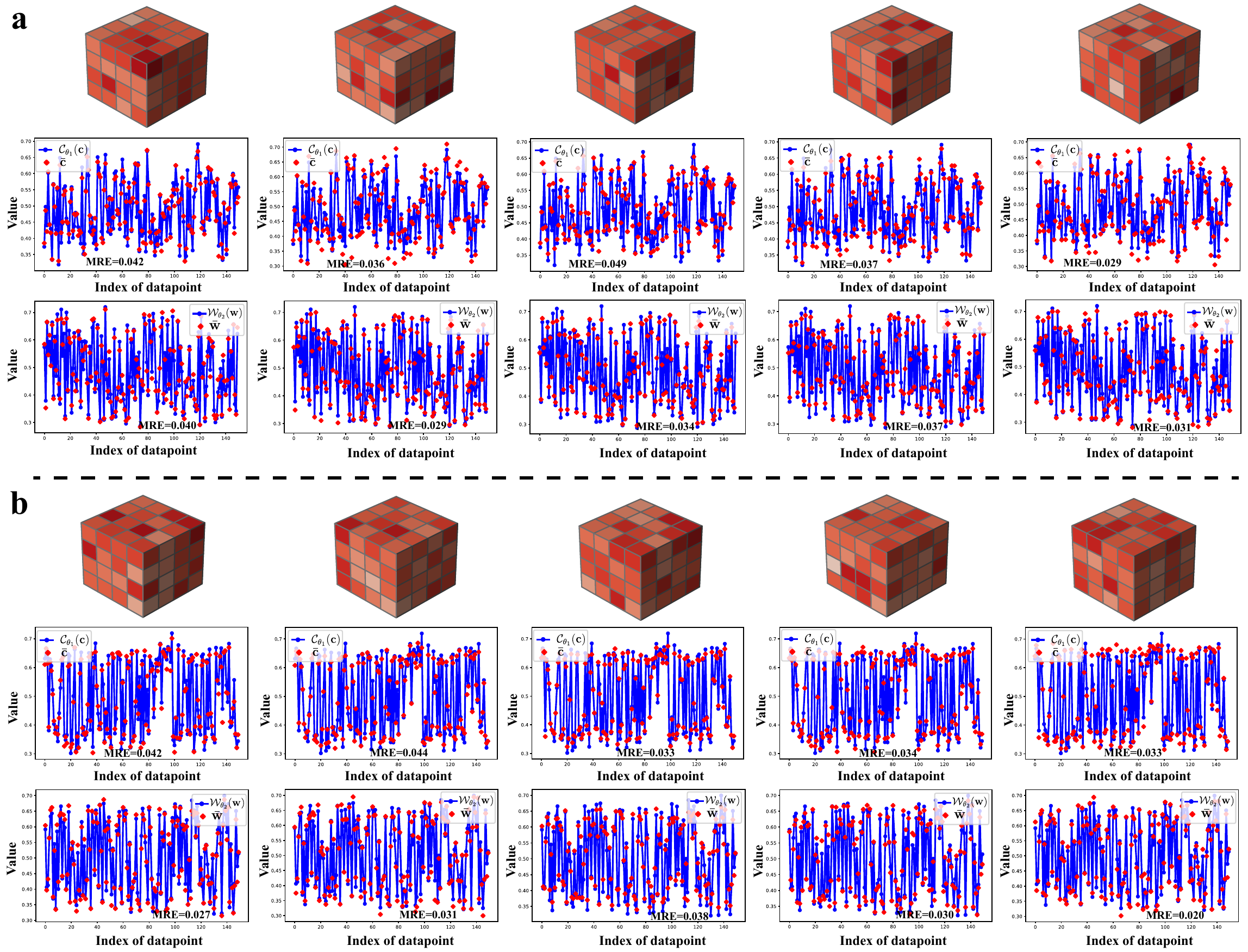}
	\caption{5 residual stress predictions for 2 concentration-averaged WC parameters pair situations}
	\label{stress prediction}
\end{figure}

\section{Gradient-based concentration optimization}

\begin{figure}[htbp]
	\centering
	\includegraphics[scale=0.35]{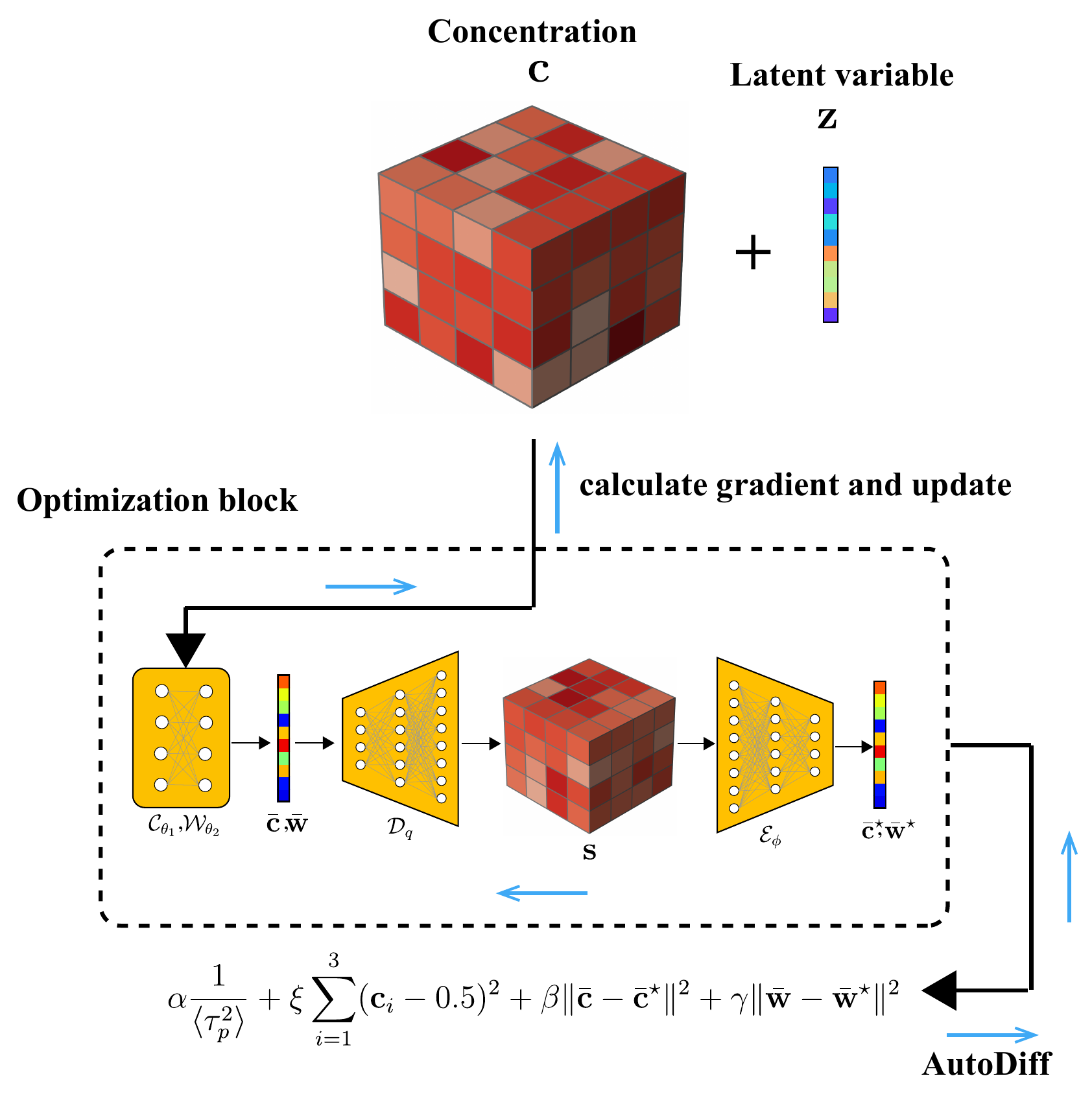}
	\caption{Demonstration of concentration design through gradient-based optimization.}
	\label{optimi_algorithm}
\end{figure}

\begin{figure}[htbp]
	\centering
	\includegraphics[scale=0.45]{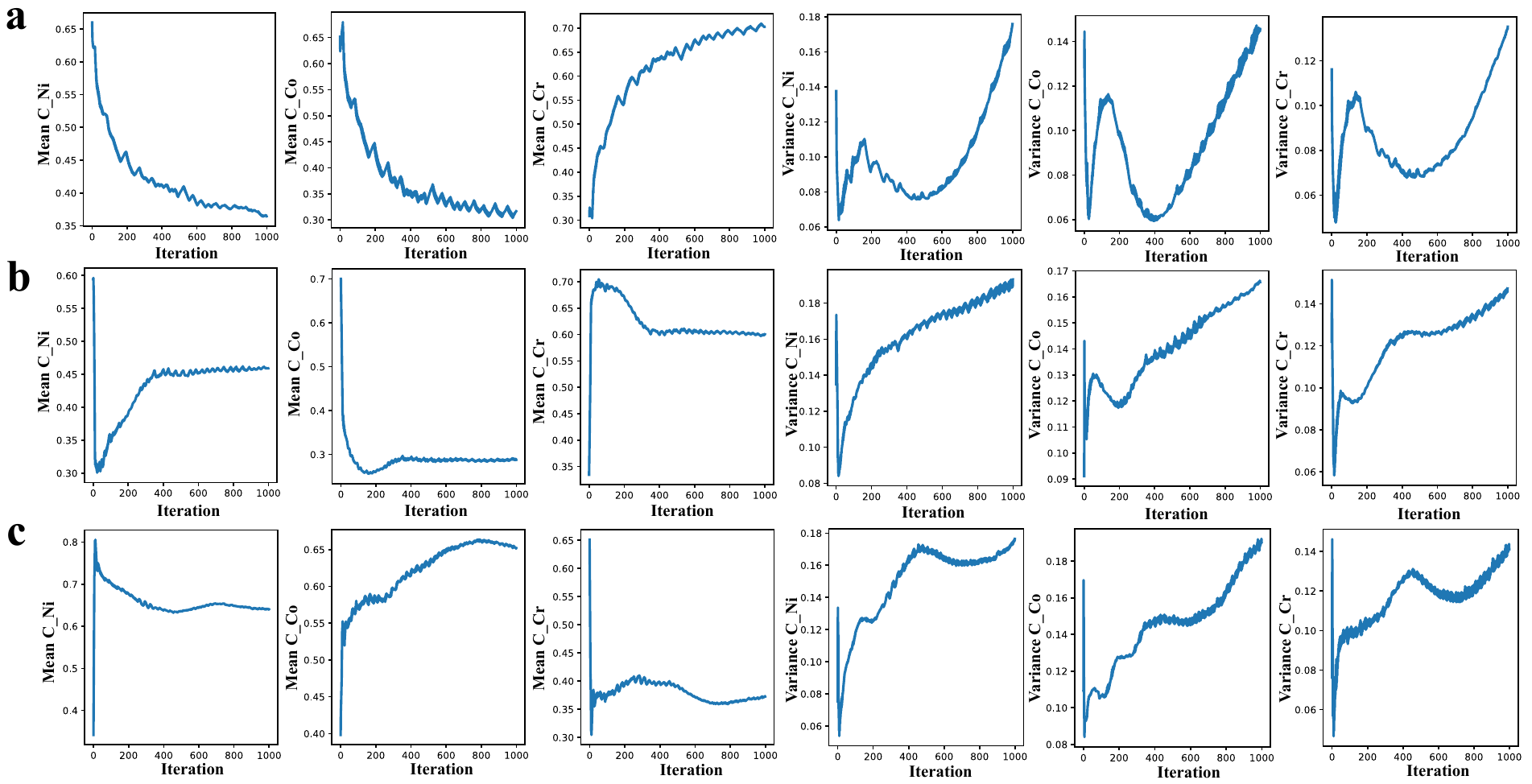}
	\caption{3 individual optimization process. First 3 columns show the concentration evaluation of 3 elements, Ni, Co, Cr; the last 3 columns show the variation of 3 elements}
	\label{3 individual ops}
\end{figure}

We give a detailed demonstration about the gradient-based concentration field optimization in Fig \ref{optimi_algorithm}. The gradient is calculated by automatic differentiation. As mentioned in Figure 7 in the main article, both 2 situation optimization results show 2 clusters. Table \ref{cluster11} lists the mean value of each element's concentration after optimizations for 2 situations. We found a coupling effect between Ni and Co element in the optimization process. Figure \ref{3 individual ops} illustrates 3 individual optimization process. We found a similar trend of concentration change of Ni and Co during the optimization process, and the variances in the concentrations of the three elements mostly display an increasing trend. These findings provide references for material design in practical applications.
\begin{table}[htbp]
\centering
\begin{tabular}{lccc}
\toprule
\textbf{cluster} & \textbf{Ni} & \textbf{Co} & \textbf{Cr} \\
\midrule
situation a cluster 1 & 0.599 & 0.602 & 0.444 \\
situation a cluster 2 & 0.416 & 0.457 & 0.545 \\
situation b cluster 1 & 0.593 & 0.588 & 0.449 \\
situation b cluster 2 & 0.429 & 0.457 & 0.549 \\
\bottomrule
\end{tabular}
\caption{The mean value of the concentration of 3 elements after optimization}
\label{cluster11}
\end{table}

\section{Additional results of application with eigenstrain}
\begin{figure}[htbp]
	\centering
	\includegraphics[scale=0.39]{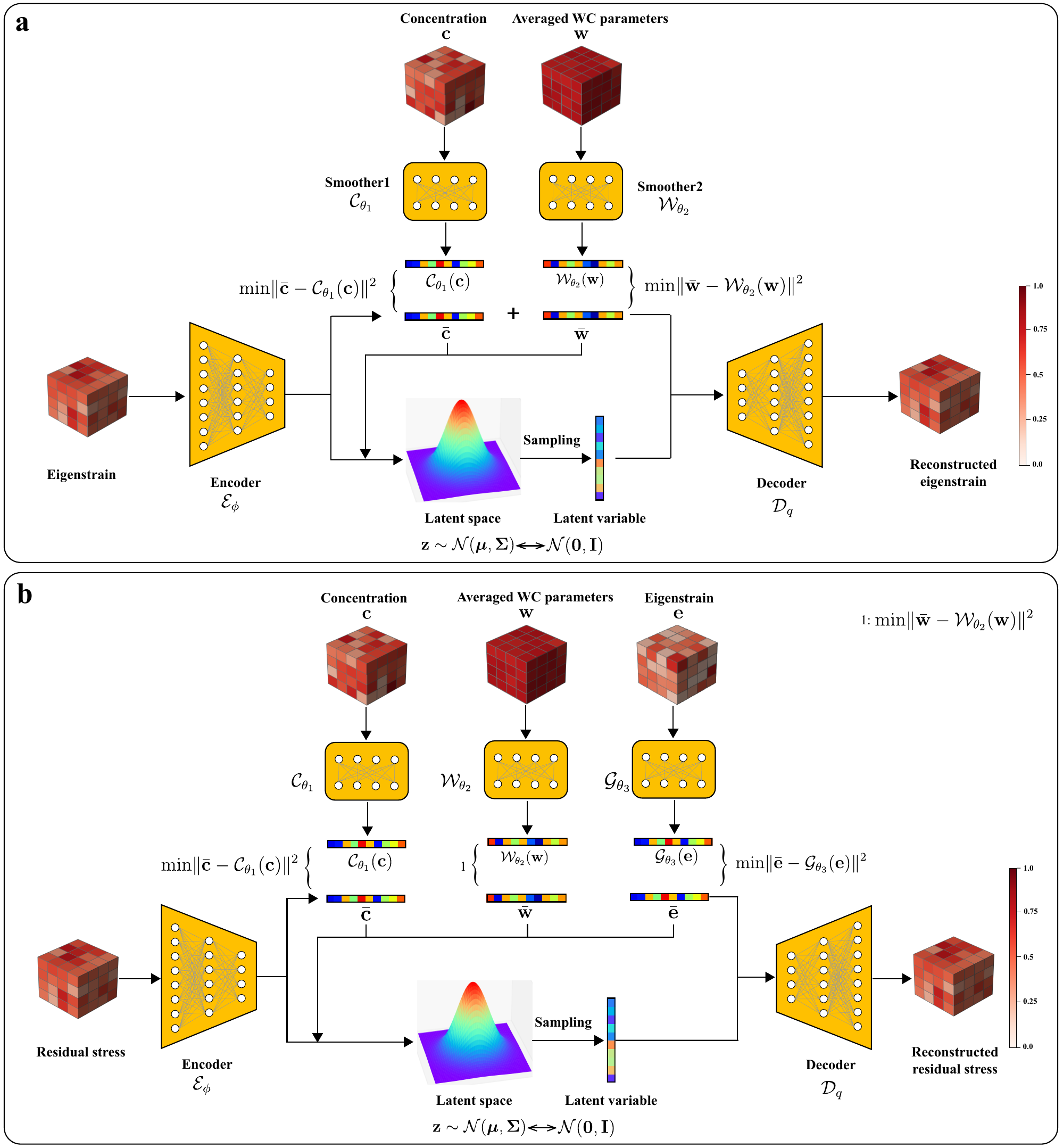}
	\caption{2 frameworks to solve other MPEA problems involving eigenstrain. a) Surrogate modeling from composition fields  to eigenstrain. b) Surrogate modeling from the composition fields and eigenstrain to residual stress.}
	\label{2 frameworks}
\end{figure}

We give the 2 frameworks involving eigenstrain in Figure \ref{2 frameworks}, both of which are slightly changed according to the general AlloyVAE framework illustrated in Figure 1 in the main article.

\end{document}